\documentclass[journal,twoside,web]{ieeecolor}
\usepackage{generic}
\usepackage{cite}
\usepackage{amsmath,amssymb,amsfonts}
\usepackage{algorithmic}
\usepackage{graphicx}
\usepackage{textcomp}
\usepackage{multirow}
\usepackage{booktabs}
\usepackage{threeparttable}

\def\BibTeX{{\rm B\kern-.05em{\sc i\kern-.025em b}\kern-.08em
    T\kern-.1667em\lower.7ex\hbox{E}\kern-.125emX}}
\begin{document}
\title{Non-contact Atrial Fibrillation Detection  from Face Videos by Learning Systolic Peaks}
\author{Zhaodong Sun, Juhani Junttila, Mikko Tulppo, Tapio Seppänen, and Xiaobai Li
\thanks{Manuscript received on November 5, 2021; This study was supported by the Academy of Finland (project 323287 and 5326291), the Finnish Work Environment Fund (Project 200414), Sigrid Juselius Foundation and Foundation for Cardiovascular Research. \emph{Corresponding author: Xiaobai Li}.}
\thanks{Zhaodong Sun, Tapio Seppänen, and Xiaobai Li are with the Center for Machine Vision and Signal Analysis, University of Oulu, FI-90014 Oulu, Finland (e-mail: zhaodong.sun@oulu.fi; tapio.seppanen@oulu.fi; xiaobai.li@oulu.fi). }
\thanks{Juhani Junttila, and Mikko Tulppo are with Research Unit of Internal Medicine, Medical Research Center Oulu, FI-90014 Oulu, Finland (e-mail: juhani.junttila@oulu.fi; mikko.tulppo@oulu.fi).}}

\maketitle

\begin{abstract}
\textit{Objective}: We propose a non-contact approach for atrial fibrillation (AF) detection from face videos. \textit{Methods}: Face videos, electrocardiography (ECG), and contact photoplethysmography (PPG) from 100 healthy subjects and 100 AF patients are recorded. Data recordings from healthy subjects are all labeled as healthy. Two cardiologists evaluated ECG recordings of patients and labeled each recording as AF, sinus rhythm (SR), or atrial flutter (AFL). We use the 3D convolutional neural network for remote PPG monitoring and propose a novel loss function (Wasserstein distance) to use the timing of systolic peaks from contact PPG as the label for our model training. Then a set of heart rate variability (HRV) features are calculated from the inter-beat intervals, and a support vector machine (SVM) classifier is trained with HRV features. \textit{Results}: Our proposed method can accurately extract systolic peaks from face videos for AF detection. The proposed method is trained with subject-independent 10-fold cross-validation with 30s video clips and tested on two tasks. 1) Classification of healthy versus AF: the accuracy, sensitivity, and specificity are 96.00\%, 95.36\%, and 96.12\%. 2) Classification of SR versus AF: the accuracy, sensitivity, and specificity are 95.23\%, 98.53\%, and 91.12\%. In addition, we also demonstrate the feasibility of non-contact AFL detection. \textit{Conclusion}: We achieve good performance of non-contact AF detection by learning systolic peaks. \textit{Significance}: non-contact AF detection can be used for self-screening of AF symptoms for suspectable populations at home or self-monitoring of AF recurrence after treatment for chronic patients.
\end{abstract}

\begin{IEEEkeywords}
Atrial Fibrillation (AF), Remote Photoplethysmography, Photoplethysmography Imaging, Wasserstein Distance, Heart Rate Variability (HRV)
\end{IEEEkeywords}

\section{Introduction}
\label{sec:introduction}

\IEEEPARstart{A}{trial} fibrillation (AF) is a common heart arrhythmia, and about 2\% of the global population is reported to have this disease \cite{authors20122012}. AF might cause stroke and heart failure, but most AF episodes are asymptomatic in the early phase. Early detection of AF episodes is essential to avoid these severe diseases. The common method for AF diagnosis is to observe electrocardiography (ECG) signals measured from the electrodes attached to the chest. However, this requires specific medical equipment and complicated operations to measure ECG signals. Cardiologists should determine the diagnosis from ECG signals, which is not practical for daily checks or long-term monitoring. Although some works \cite{fukunami1991detection, mehta2009detection,guidera1993signal,couceiro2008detection,tateno2001automatic,dash2009automatic,islam2016rhythm,islam2019robust,couceiro2008detection,shashikumar2017deep,fan2018multiscaled,pourbabaee2018deep} designed automatic AF detection algorithms from ECG and achieve a high accuracy, the acquisition of ECG signals is complicated and not convenient for daily use. Some works \cite{poh2018diagnostic, kwon2019deep, eerikainen2019detecting,lee2012atrial, krivoshei2017smart, bonomi2018atrial, dorr2019watch, corino2017detection, valiaho2019wrist} propose to use contact photoplethysmography (PPG) from pulse oximeters, smartwatches or phone cameras on fingers for more convenient and accessible AF detection. However, all of the ECG and contact PPG methods require the sensors attached to the body/skin, which might cause irritation and hygiene issues. In addition, specific biomedical equipment such as ECG sensors or pulse oximeters also decrease the accessibility of these methods to more populations.

Compared with ECG and contact PPG, remote PPG\footnote{Remote PPG is used to be literally distinguished from contact PPG. There are other terms equivalent to remote PPG such as photoplethysmography imaging \cite{wu2000photoplethysmography}, video photoplethysmography \cite{iozzia2016relationships}, and videoplethysmography \cite{couderc2015detection}} provides a non-contact way to monitor cardiac signals by using a camera. Remote PPG is measured from the face color change induced by the blood volume change in face videos \cite{wu2000photoplethysmography, verkruysse2008remote}, and can reveal cardiac rhythms for AF detection. As cameras are ubiquitous nowadays, this might lead to potential solutions or products for convenient non-contact AF detection. There are several situations where AF detection from face videos is preferred. First, AF detection from face videos can be used for telemedicine during video conferences without specific medical instruments. Second, contact AF detection is not applicable to people with skin burns or other skin diseases on the measurement locations such as fingers or chest. It has been reported that 486,000 people in the United States were burned to receive medical treatment in the year 2011 \cite{us_burn}. Third, cameras are ubiquitous and accessible such as in smartphones or laptops, and much cheaper than the smartwatches with ECG/PPG sensors. Finally, as people spend a long time on using smart phones or laptops every day, the AF detection from face videos can be integrated to smart phones or laptops when people are using these devices. 

Recent studies \cite{couderc2015detection,li2018obf,yu2019remote,shi2019atrial,yan2018contact, corino2020simple} have made some contributions to non-contact AF detection by remote PPG. Most studies focus on heart rate variability (HRV) features derived from systolic peaks in remote PPG for AF detection and achieved promising results. The problem is that remote PPG signals are subtle compared to noises, and systolic peaks of AF patients have low amplitudes when the heart rate is high \cite{valiaho2019wrist}. Therefore, it is challenging to develop a method that can measure systolic peaks accurately enough for AF detection. Therefore, we develop a method to accurately measure systolic peaks from face videos to facilitate non-contact AF detection and test the effectiveness of the method on our large-scale dataset.

Our main contributions are listed below:
\begin{itemize}
\item We develop a deep learning-based remote PPG algorithm by using the timing of the systolic peaks as the ground truth and demonstrate that learning systolic peaks from face videos can facilitate non-contact AF detection.
\item We record the full version of Oulu Bio-face (OBF) dataset \cite{li2018obf} for non-contact AF detection, including videos from 100 healthy subjects and 100 AF patients. Two cardiologists evaluated ECG recordings of patients and labeled each recording as AF, sinus rhythm (SR), or atrial flutter (AFL).
\item We test two types of AF detection models for the self-screening of healthy subjects and patients, respectively, and achieve high accuracy. The classification between healthy subjects and patients with AF has an accuracy of 96.00\%. The classification between patients with SR and patients with AF has an accuracy of 95.23\%. Additional experiments also demonstrate the feasibility of non-contact AFL detection.
\end{itemize}

\section{Related Works}
\subsection{Contact Methods for AF detection}

The AF detection from ECG can be based on 1) P-wave detection \cite{fukunami1991detection, mehta2009detection, guidera1993signal,couceiro2008detection}, 2) R-R interval variability \cite{tateno2001automatic, dash2009automatic, islam2016rhythm, islam2019robust,couceiro2008detection}, and 3) deep learning (DL)-based methods \cite{shashikumar2017deep, fan2018multiscaled, pourbabaee2018deep}. P waves are not prominent features and are often inﬂuenced by artifacts, while R peaks have higher amplitudes that can be easily detected. AF patients have a larger R-R interval variability, which means the cardiac rhythm of AF patients shows larger irregularity, so the features from R-R intervals can be a good indicator for AF detection. Some traditional methods \cite{tateno2001automatic,dash2009automatic,islam2016rhythm} extracted features such as entropy \cite{islam2016rhythm} or HRV features \cite{dash2009automatic} from R-R intervals and simply use a threshold to do AF classification. Some methods \cite{islam2017atrial,andersen2019deep,islam2019robust} applied machine learning methods such as support vector machine (SVM) \cite{islam2017atrial}, convolutional neural network \cite{andersen2019deep} or neighborhood component analysis \cite{islam2019robust} to R-R intervals for AF detection and achieved higher performance than traditional methods. However, it is difficult for rhythm-based methods using R-R intervals to distinguish AF from other arrhythmias, while the morphology-based methods such as P-wave absence can avoid false-positive errors in AF detection \cite{butkuviene2021considerations}. Deep learning-based methods \cite{shashikumar2017deep, fan2018multiscaled, pourbabaee2018deep} can be directly applied to ECG signals and achieve good AF detection performance with large datasets, but they are sensitive to changes in ECG morphology and are unclear how they generalize to unseen data \cite{butkuviene2021considerations}.

Although ECG is a gold standard for AF diagnosis, some studies found contact PPG \cite{poh2018diagnostic, kwon2019deep, eerikainen2019detecting,lee2012atrial, krivoshei2017smart, bonomi2018atrial, dorr2019watch, corino2017detection, valiaho2019wrist} measured from ﬁngertips and ballistocardiograph (BCG) \cite{bruser2012automatic,hurnanen2016automated,lahdenoja2017atrial} can also be used for AF detection. PPG and BCG signals do not have the P-wave in ECG, so the systolic peaks in PPG and BCG are the only clue for AF detection. Similar to the R-peaks in ECG, the systolic peaks are also an indicator of cardiac rhythm and the peak-peak intervals can also be used to derive the HRV features. For the daily use and better accessibility, some work proposed to use the accelerometer and gyroscope of a smartphone on the chest to get the BCG signal for AF detection [10], while others proposed to use the camera of a smartphone [6] on the ﬁngertips or smart watches \cite{bonomi2018atrial, dorr2019watch, corino2017detection, valiaho2019wrist} to get the contact PPG signals for AF detection. The AF detection from contact PPG and BCG provides other alternatives to ECG signals and is more convenient for daily health monitor.

\subsection{Remote PPG for Non-contact AF detection}

Recent studies \cite{couderc2015detection, yan2018contact, shi2019atrial, corino2020simple} showed that remote PPG from face videos can be used for non-contact AF detection. Couderc et al. \cite{couderc2015detection} first used HRV features from remote PPG for AF detection on a small dataset of 11 patients, which demonstrates the feasibility of AF detection from face videos. Yan et al. \cite{yan2018contact} collected a dataset including 217 patients among which 75 patients showed AF. They used a pre-trained SVM with features from autocorrelation analysis and achieved 95.4\% accuracy for classifying AF patients vs. non-AF patients. Shi et al. \cite{shi2019atrial} proposed a HRV feature fusion approach by combining three remote PPG algorithms \cite{wang2016algorithmic,li2018obf,feng2014motion} for AF detection. The method was tested on the previous version of OBF dataset \cite{li2018obf}, including 30 AF patients and 100 healthy subjects, and achieved an accuracy of 92.56\%. Corino et al. \cite{corino2020simple} used a simple model with three features to do AF detection from remote PPG on 68 patients and achieved 87\% AF detection accuracy.

There are one or more of the following limitations in the previous works. 1) Their AF detection accuracy is limited, as the systolic peak information was not exploited for training remote PPG algorithms. We propose to utilize systolic peaks to train our remote PPG algorithm, and the reasons are in two aspects. First, studies \cite{lee2012atrial, eerikainen2019detecting, couderc2015detection, li2018obf, yu2019remote, shi2019atrial} showed that HRV features computed using systolic peaks are sufficient for AF detection, so the waveform is not needed. Second, the systolic peaks are the most prominent and reliable feature in remote PPG, while other waveform features are subtle and easily contaminated by noises. 2) Most previous studies were tested on limited AF cases, and a larger scale of AF data is needed to make a reliable evaluation of the methods. 3) The low diversity of previous datasets also limits their application scope. Previous studies mainly focused on classifying healthy subjects versus before-treatment AF patients. It is also essential to compare before-treatment (AF) and after-treatment (sinus rhythm, SR) of the same patients, which was not concerned in any previous study yet. 4) Previous studies (except \cite{li2018obf}) did not compare the non-contact AF detection performance with results from ECG or contact PPG. The AF detection results from ECG or contact PPG can be a performance upper bound from which we can see the gap between non-contact and contact AF detection.


\section{Method}

In this section, we will introduce the 3-dimensional convolutional neural network (3DCNN) model for remote PPG monitoring and how to select a loss function to train our model with the timing of systolic peaks to facilitate AF detection. We will also explain using the HRV features derived from systolic peaks for AF detection.

\begin{figure*}[hbt!]
\begin{minipage}[b]{\linewidth}
  \centering
  \centerline{\includegraphics[width=\linewidth]{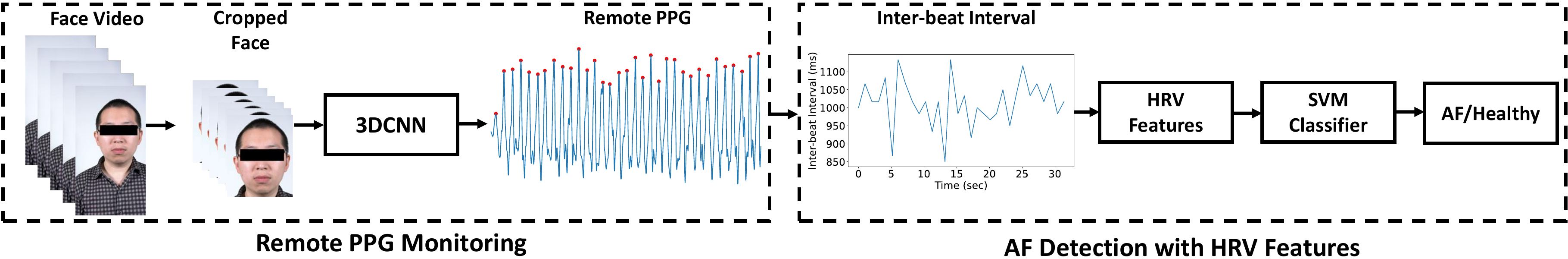}}
\end{minipage}
\caption{The general framework of non-contact AF detection. The red dots in the remote PPG are the systolic peaks. 3DCNN = three dimensional convolutional neural network, PPG = photoplethysmography, AF = atrial fibrillation, HRV = heart rate variability, SVM = support vector machine.}
\label{fig:framework}
\end{figure*}

\begin{figure}[hbt!]
\centering
\begin{minipage}[b]{0.8\linewidth}
  \centering
  \centerline{\includegraphics[width=\linewidth]{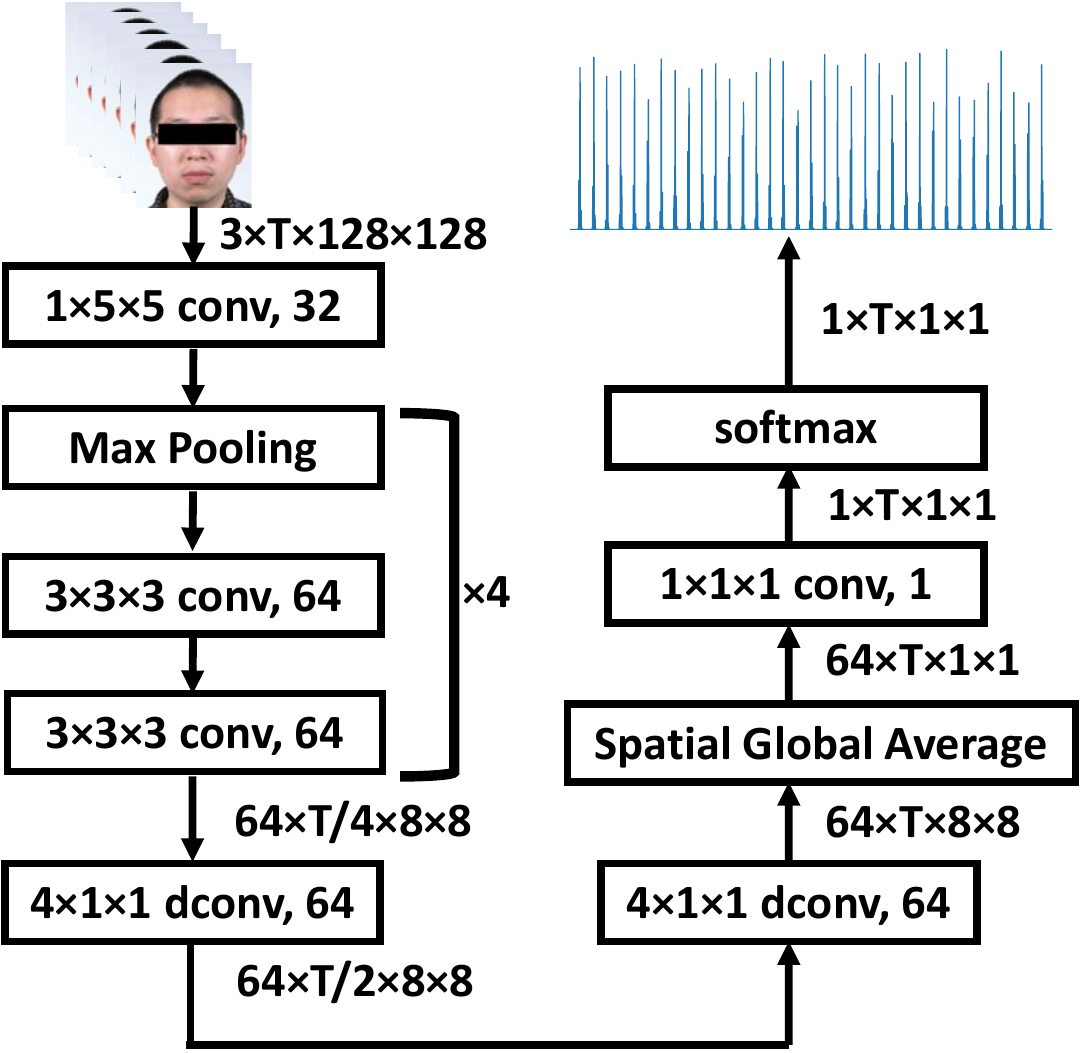}}
\end{minipage}
\caption{Three-dimensional convolutional neural network (3DCNN) architecture. The input is the image sequence of the cropped face. The output is the systolic peak signal. "3x3x3 Conv, 64" means the 3D convolution operation with filter size 3x3x3 and the output channel is 64. "dconv" is 3D transposed convolution, which works as upsampling. There is also a batchnorm layer and a ReLU activation following each convolutional block. There are 4 max pooling layers. All of them downsample the spatial dimensions, and two of them only downsample the temporal dimension.}
\label{fig:3dcnn}
\end{figure}


\subsection{Remote PPG Monitoring}

\subsubsection{Preprocessing}
We first need to crop the face region from the face video. The face region is obtained from the landmarks generated from OpenFace \cite{baltrusaitis2018openface}. We first get the minimum and maximum horizontal and vertical coordinates of the landmarks to locate the central face point. The bounding box size is 1.2 times the range of vertical coordinates of landmarks and is fixed for each video. After getting the central point and the size of the bounding box, we can crop the face region from each frame as shown in Fig. \ref{fig:framework}. The cropped face is resized to $128 \times 128$.

\subsubsection{3DCNN for Remote PPG Monitoring}
The cropped face is fed into the 3DCNN model \cite{yu2019remote} for remote PPG monitoring. 3DCNN uses the 3D kernels to do the convolution on the video along the width, height, and time axis. We will use the similar 3DCNN architecture as \cite{yu2019remote} with one modification that will be illustrated in the next part. The model is shown in Fig. \ref{fig:3dcnn}. The input for 3DCNN is the cropped face video clip $x \in \mathbb{R}^{C \times H \times W \times T}$ where $C$ is the number of color channels, $H$ is the video height, $W$ is the video width, and $T$ is the video time length. The output remote PPG is $p_r = G_{\theta}(x) \in \mathbb{R}^{T}$. The input video is a 4D video signal, while the network eliminates the height, width, and color dimensions and converts the 4D signal into a 1D remote PPG signal. To train this network, we need to minimize a loss function. The loss function measures the distance between the ground truth, which is the contact PPG $p_c(t)$, and the network output, which is the remote PPG $p_r(t)$. A previous method \cite{yu2019remote} used negative Pearson correlation as the loss function with the contact PPG as the ground truth. This loss function encourages the remote PPG signals and contact PPG signals to have similar morphology. However, this loss function also encourages the model to learn redundant information from the contact PPG, such as diastolic peaks or artifacts.


\subsubsection{Binary Systolic Peaks for 3DCNN Training}

\begin{figure}[hbt!]

\begin{minipage}[b]{\linewidth}
  \centering
  \centerline{\includegraphics[width=\linewidth]{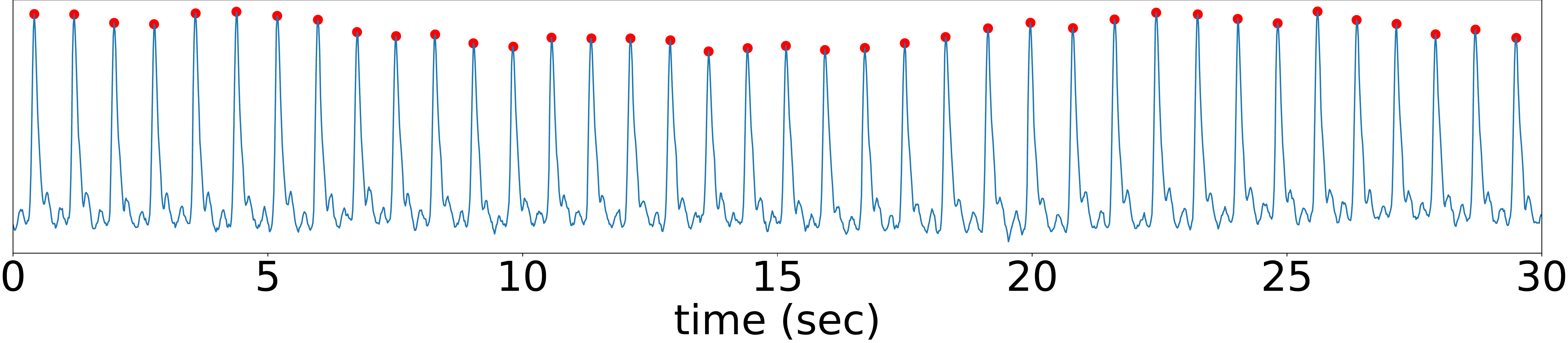}}
  \centerline{(a) Contact PPG from a healthy subject}
\end{minipage}

\vfill

\begin{minipage}[b]{\linewidth}
  \centering
  \centerline{\includegraphics[width=\linewidth]{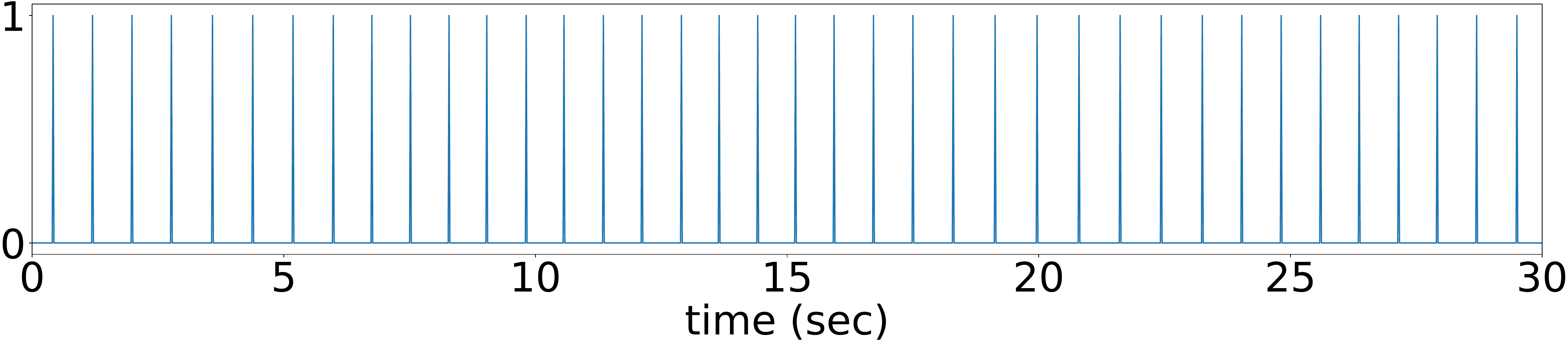}}
  \centerline{(b) Binary systolic peaks from a healthy subject}
\end{minipage}

\vfill

\begin{minipage}[b]{\linewidth}
  \centering
  \centerline{\includegraphics[width=\linewidth]{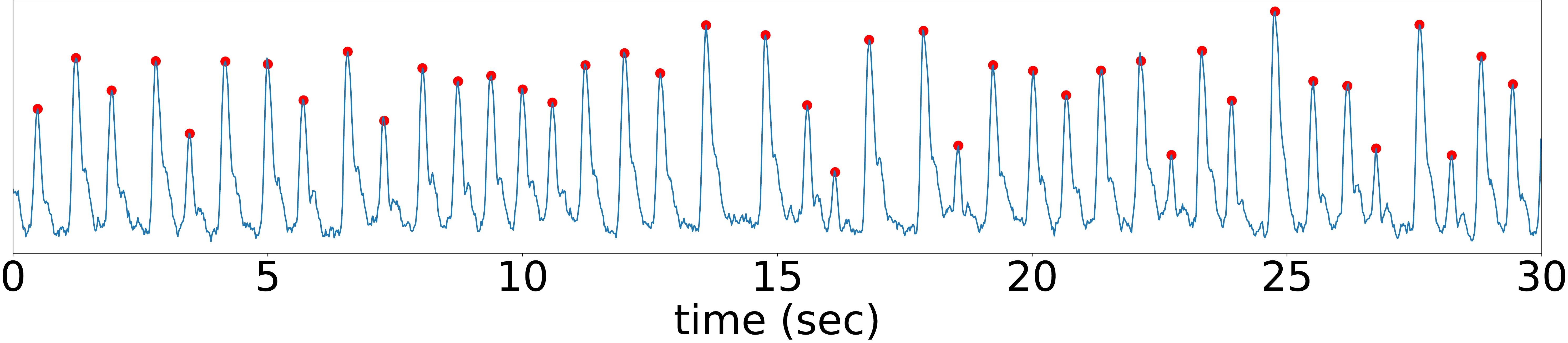}}
  \centerline{(c) Contact PPG from a patient with AF}
\end{minipage}

\vfill

\begin{minipage}[b]{\linewidth}
  \centering
  \centerline{\includegraphics[width=\linewidth]{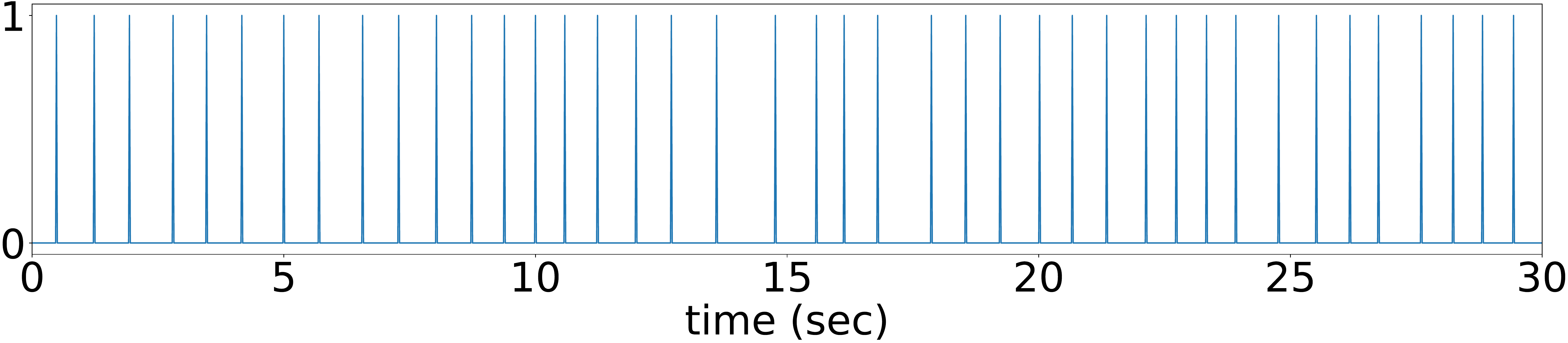}}
  \centerline{(d) Binary systolic peaks from a patient with AF}
\end{minipage}

\caption{Contact PPG (with systolic peaks marked with red dots) and corresponding binary systolic peaks. Binary systolic peaks (b and d) are extracted from the contact PPG signals (a and c). The PPG signal from the AF patient contains complex patterns such as irregular heartbeat rhythms and different systolic peak amplitudes. PPG = photoplethysmography, AF = atrial fibrillation.}
\label{fig:cppg_healthy_pa}
\end{figure}

Using contact PPG as the ground truth to train the network is not the best. The systolic peak timing in the contact PPG is a better ground truth. There are two reasons. First, \cite{mcduff2014remote} demonstrated that the waveform morphology of contact PPG signals is not entirely consistent with remote PPG due to their different measurement locations. This means that contact PPG is not the actual ground truth for each face video due to the inconsistency of contact and remote PPG. However, \cite{mcduff2014remote} also showed that the timing of systolic peaks is the consistent information between contact and remote PPG. Second, contact PPG from patients usually contain systolic peaks with much lower amplitude than that from healthy subjects, as shown in Fig. \ref{fig:cppg_healthy_pa}. Väliaho et al. \cite{valiaho2019wrist} also reported this phenomenon and explained that a higher heart rate in AF causes smaller stroke blood volume and lower systolic peak amplitudes. If we use the contact PPG as the ground truth for model training, the model might learn some redundant information and ignore some systolic peaks with low amplitude. Therefore, we will convert our contact PPG into binary systolic peaks as shown in Fig. \ref{fig:cppg_healthy_pa}. The binary systolic peaks only have zeros and ones, which means it only keeps the timing of the systolic peaks and removes other redundant information. This kind of ground truth can encourage the model only to learn the timing of systolic peaks, which is also the only information used for AF detection.

The next question is how to use the binary systolic peaks for our model training. We can regard our model output $p_r = G_{\theta}(x)$ and our binary systolic peak label $s$ as two probability distributions so that we can use some probability distribution distances as our loss function. The reason is that probability distribution distances can compare a binary pulse signal with a continuous smooth signal. Our systolic peak signal is just a series of binary signals, and our model output is the continuous smooth signal, so probability distribution distances might be a good choice. Previous used negative Pearson correlation loss function can handle continuous values in contact PPG but cannot work with binary values in binary systolic peak signals.  

To satisfy the requirement of probability distributions, we should ensure that the sum of a signal is one and the signal is non-negative. We can normalize the binary systolic peak as $\bar{s} = s / \Sigma_i s_i$. Since the softmax layer can make sure the model output is non-negative and the sum is one, we add a softmax layer to the output of the network $G_{\theta}(x)$ and the new network is $H_{\theta}(x) = \text{softmax}(G_{\theta}(x))$. Therefore, we have our new model output $\bar{p}_r = H_{\theta}(x)$. Our new loss function $l(\bar{s}, \bar{p}_r)$ is defined to measure the distance between normalized systolic peak signal $\bar{s}$ and model output $\bar{p}_r$. We use Wasserstein distance as our loss function $l$ for learning the systolic peaks. The reason why Wasserstein distance is used as the loss function and the full analysis about the loss function selection is described in the following.

\subsection{Loss Function Selection for Learning Systolic Peaks}
\label{sec:loss_select}

\subsubsection{Candidate loss functions}
\label{sec:loss}

For the proposed 3DCNN model, both the model output signal $\bar{p}_r$ and the normalized binary systolic peaks $\bar{s}$ can be regarded as probability distributions. Therefore, we can use some probability distribution distances to measure their similarity. These probability distribution distances can be used as loss functions for the model training. During training, the loss function is minimized so that our model output signal is well aligned with the systolic peak signals. In previous work \cite{cazelles2020wasserstein}, probability distribution distances were used for comparing two time series in the power spectral density (PSD) domain. However, the accurate representation of systolic peaks can only be in the time domain. Therefore, we directly use the probability distribution distances in the time domain. There are some options for probability distribution distances, such as squared Euclidean distance, Kullback-Leibler (KL) divergence, Jensen–Shannon (JS) divergence, and Wasserstein distance.

\textbf{Squared Euclidean Distance (SED):}
Squared Euclidean distance is a straightforward way to compare two time series. It measures the Euclidean distance between two probability distributions. It is also widely used as a loss function for deep learning models. It is defined as
\begin{equation}
l_{SED}(\bar{p}_r, \bar{s}) = \parallel \bar{p}_r - \bar{s} \parallel_2^2
\end{equation}

\textbf{Kullback-Leibler (KL) Divergence:}
The KL divergence is another probability distribution distance widely used as the loss function in classification tasks. In the classification tasks, it measures the distance between the true classification distribution and the predicted classification distribution. KL divergence is defined as
\begin{equation}
l_{KL}(\bar{p}_r, \bar{s})  = \text{KL}(\bar{s} \parallel \bar{p}_r) = \sum_{t=1}^T \bar{s}(t) \log \Big( \frac{\bar{s}(t)}{\bar{p}_r(t)} \Big)
\end{equation}

\textbf{Jensen–Shannon (JS) Divergence:}
JS divergence is based on KL divergence. Compared with KL divergence, JS divergence is symmetric and bounded between 0 and 1. It is defined as
\begin{equation}
l_{JS}(\bar{p}_r, \bar{s}) = \frac{1}{2}\text{KL}(\bar{p}_r \parallel \frac{\bar{p}_r+\bar{s}}{2}) + \frac{1}{2}\text{KL}(\bar{s} \parallel \frac{\bar{p}_r+\bar{s}}{2})
\end{equation}

\begin{figure}[hbt!]

\begin{minipage}[b]{\linewidth}
\begin{minipage}[b]{0.48\linewidth}
  \centering
  \centerline{\includegraphics[width=\linewidth]{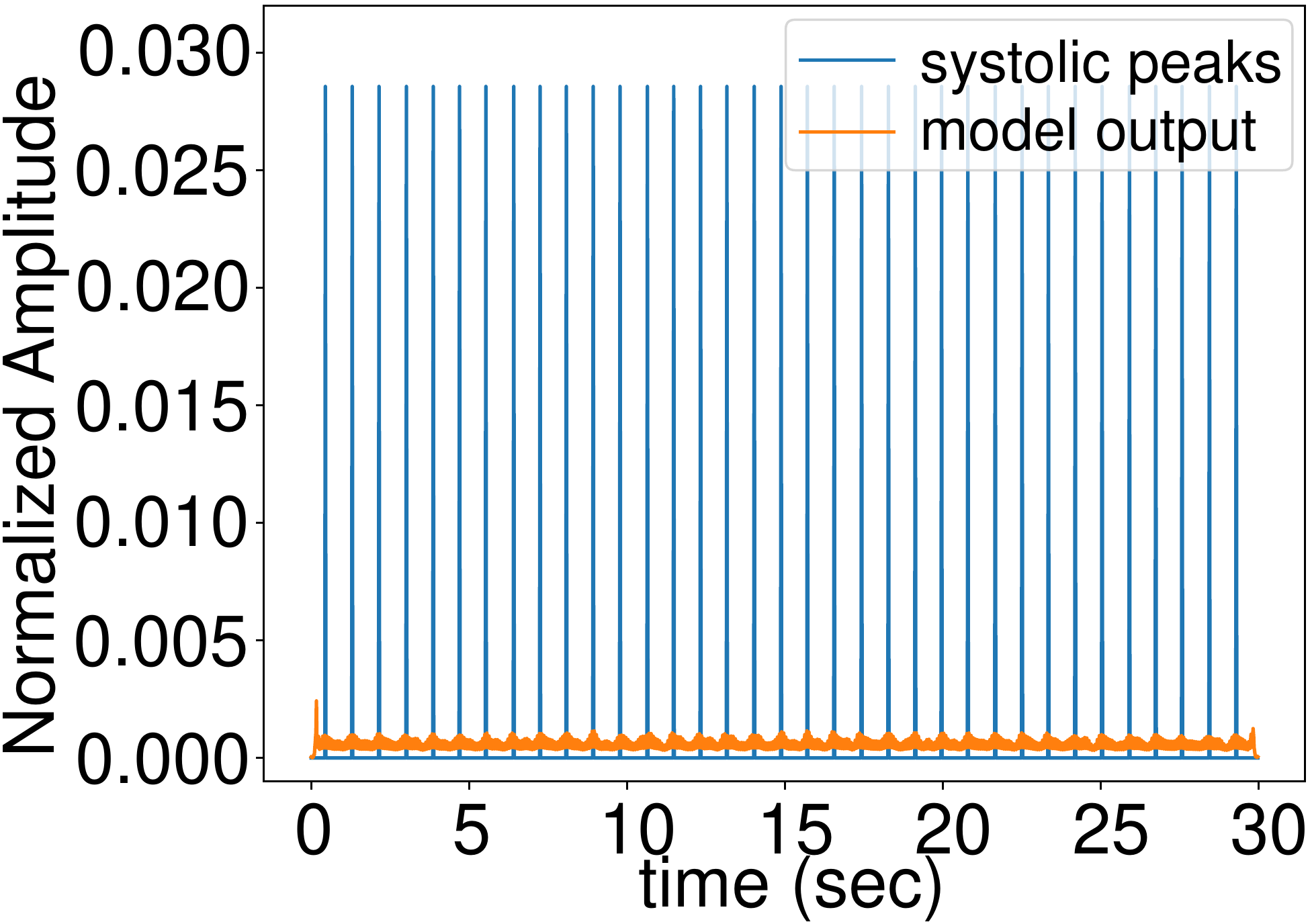}}
\end{minipage}
\begin{minipage}[b]{0.48\linewidth}
  \centering
  \centerline{\includegraphics[width=\linewidth]{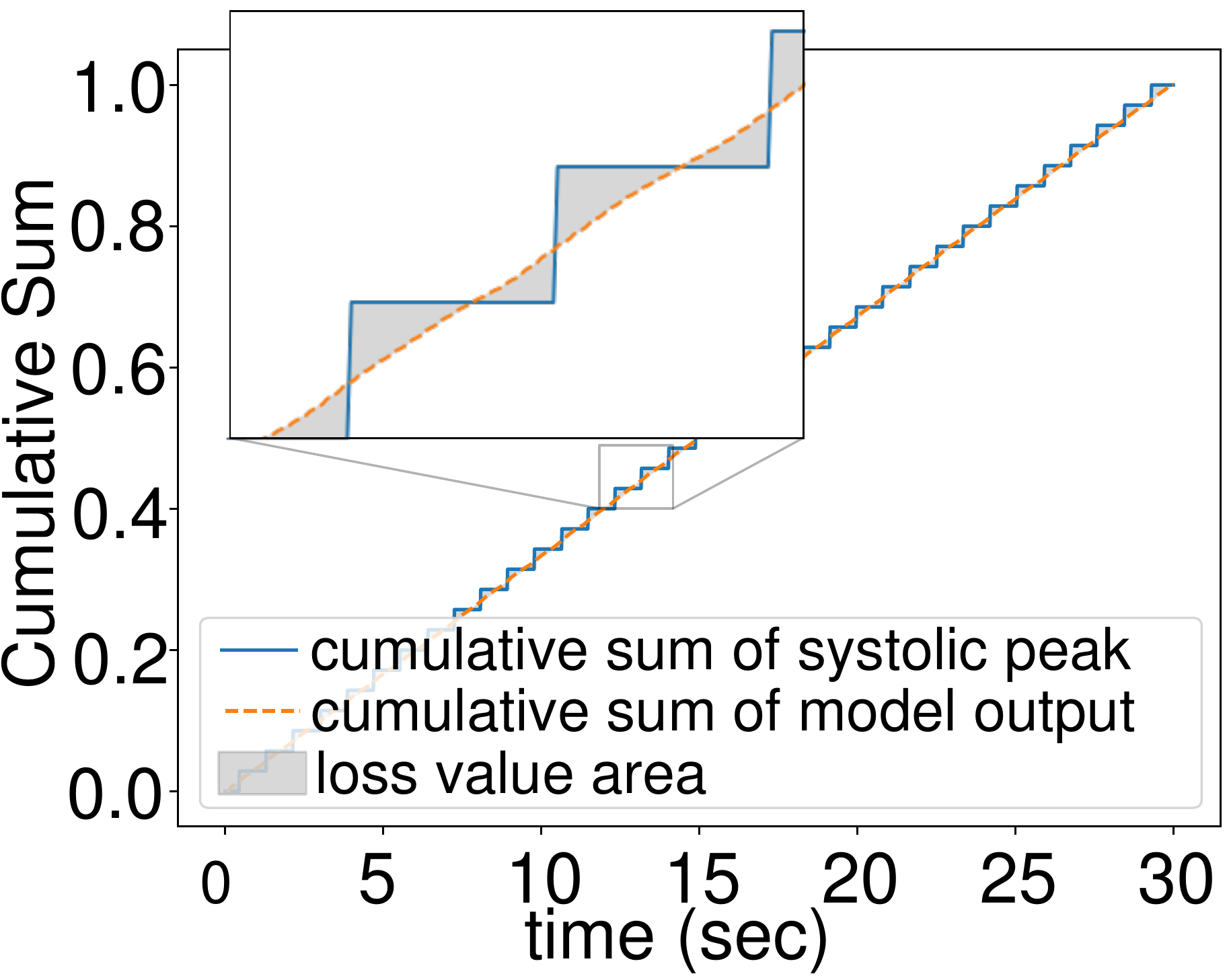}}
\end{minipage}
\centerline{(a) epoch 1}
\end{minipage}

\begin{minipage}[b]{\linewidth}
\begin{minipage}[b]{0.48\linewidth}
  \centering
  \centerline{\includegraphics[width=\linewidth]{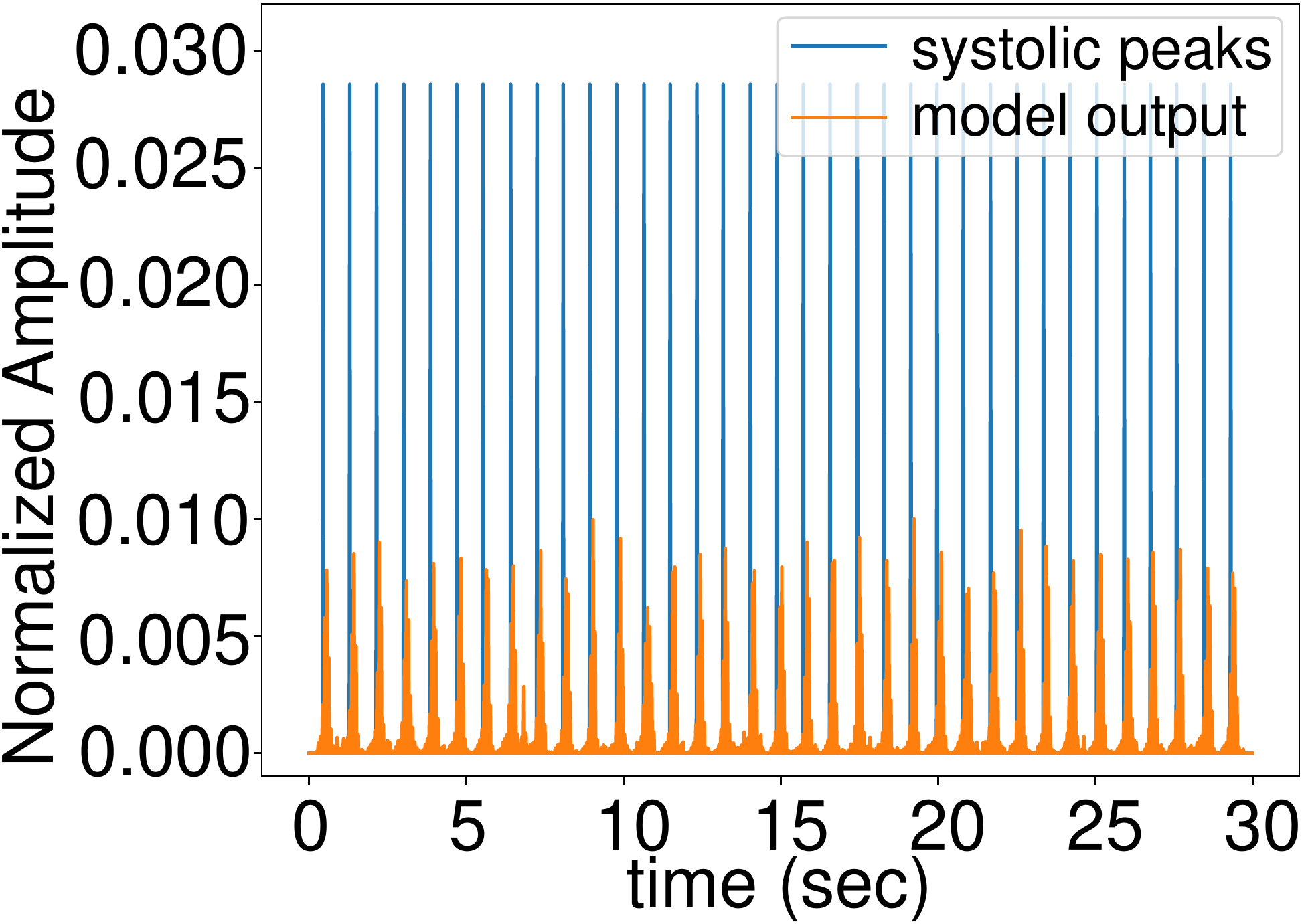}}
\end{minipage}
\begin{minipage}[b]{0.48\linewidth}
  \centering
  \centerline{\includegraphics[width=\linewidth]{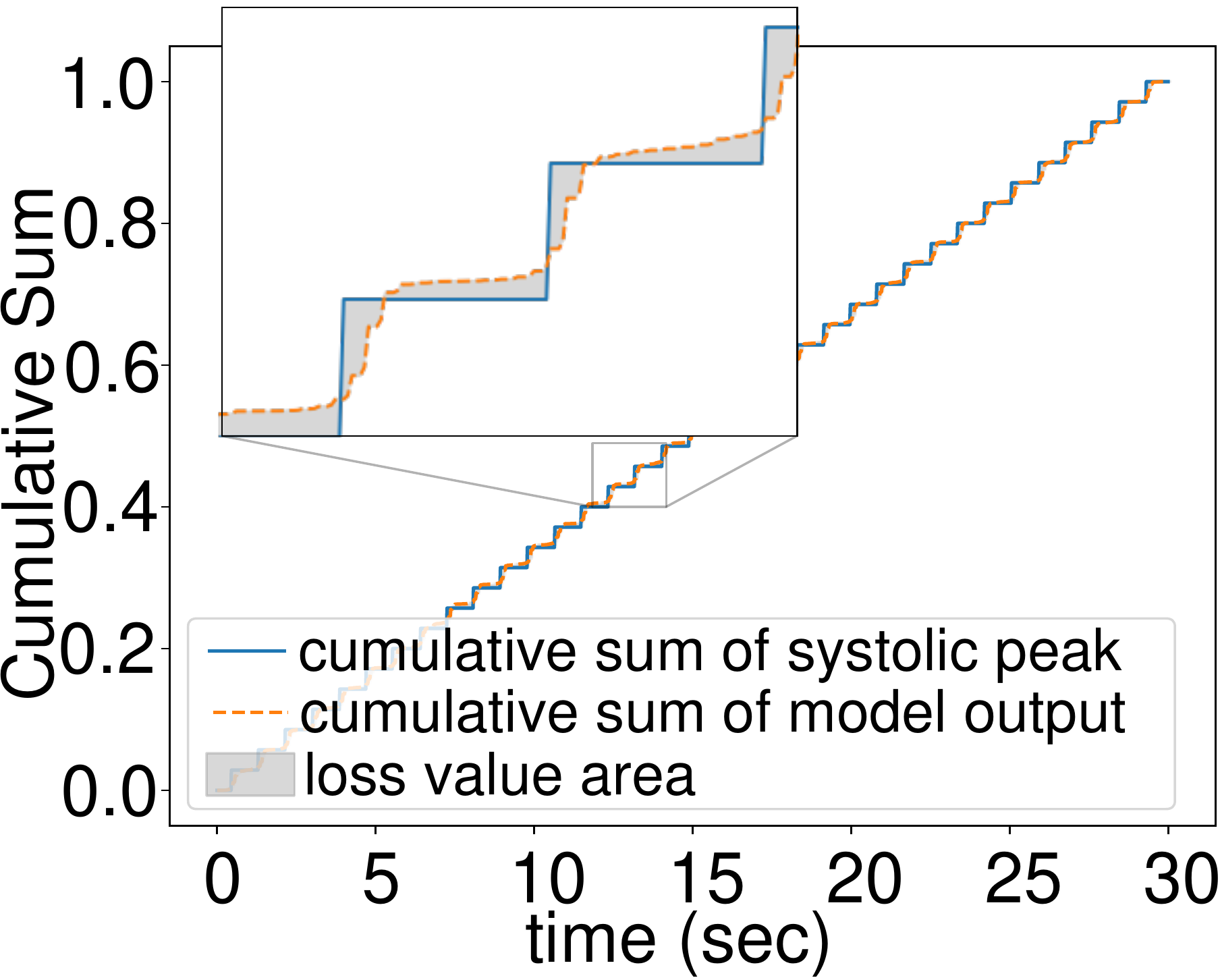}}
\end{minipage}
\centerline{(b) epoch 15}
\end{minipage}

\begin{minipage}[b]{\linewidth}
\begin{minipage}[b]{0.48\linewidth}
  \centering
  \centerline{\includegraphics[width=\linewidth]{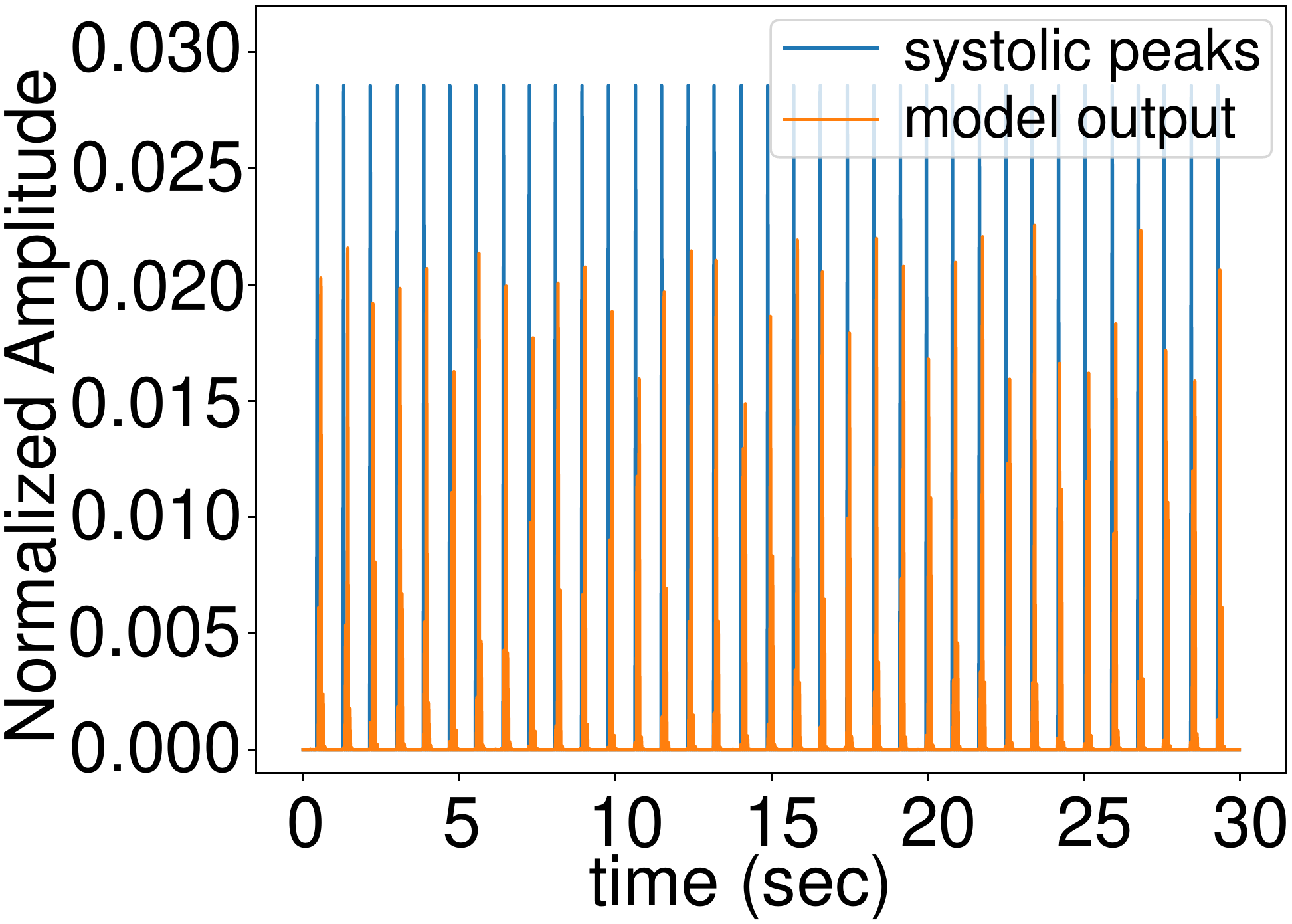}}
\end{minipage}
\begin{minipage}[b]{0.48\linewidth}
  \centering
  \centerline{\includegraphics[width=\linewidth]{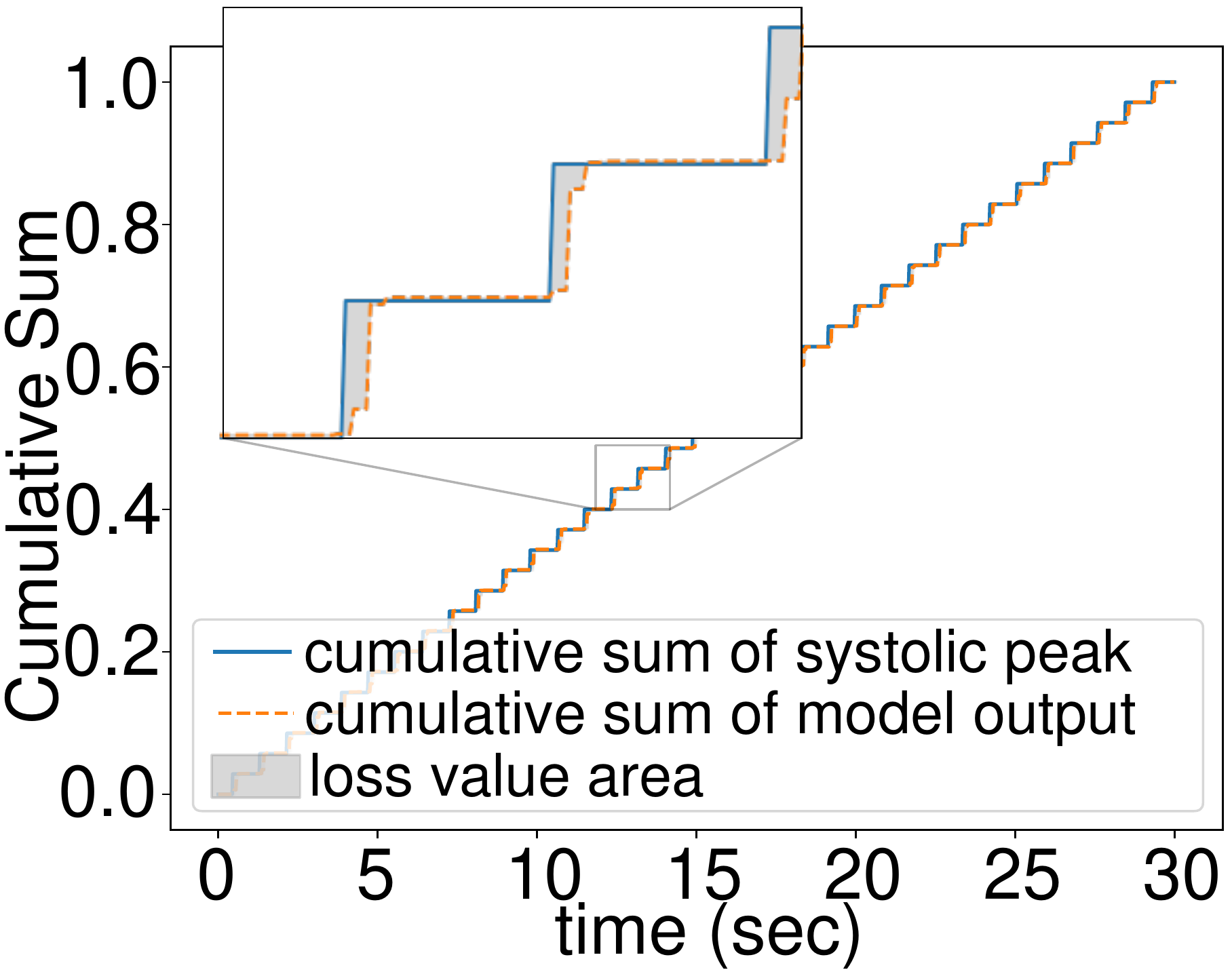}}
\end{minipage}
\centerline{(c) epoch 45}
\end{minipage}

\caption{Illustration of how Wasserstein distance works. The first, second, and third rows are at epochs 1, 15, and 45 of model training. The first column shows the ground truth systolic peaks and our model output. The second column shows the cumulative sum of the two signals in the first column and also shows the zoom-in figures. The gray area is the Wasserstein distance value. During model training, we intend to minimize this gray area so that the model output is similar to the systolic peak signal. The decreasing of the area makes the peak signal from the model sharper and more accurately aligned with the true systolic peaks.}

\label{fig:ws_work}
\end{figure}

\textbf{Wasserstein Distance:}
Wasserstein distance is based on optimal transport. It finds the minimum cost to move the mass of one probability distribution to turn the probability distribution into another. It is defined as
\begin{equation}
l_{WS} (\bar{p}_r, \bar{s}) = \min_{\pi \in \Pi(\bar{p}_r, \bar{s})} \mathbb{E}_{(x,y) \sim \pi } [\parallel x-y \parallel]
\end{equation}
where $\Pi(\bar{p}_r, \bar{s})$ is a set containing all joint probability distributions with marginal distribution $\bar{p}_r$ and $\bar{s}$. Wasserstein distance finds the joint probability distribution (transport plan) $\pi$ so that the movement cost is minimum. $x$ and $y$ are two random variables following the joint distribution $\pi$. However, the formula above has a min operation and cannot be directly used as a loss function. Since our signals $\bar{p}_r$ and $\bar{s}$ are one-dimensional, the Wasserstein distance has a closed-form in the one-dimensional case, as shown below.
\begin{equation}
l_{WS} (\bar{p}_r, \bar{s}) = \int_0^1 |F_{\bar{p}_r}^- (u) - F_{\bar{s}}^-(u)| du = \sum_{t=1}^T |F_{\bar{p}_r} (t) - F_{\bar{s}}(t)|
\label{eq:final_ws}
\end{equation}
where $F_{\bar{p}_r}$ is the cumulative sum of $\bar{p}_r$, which is also the cumulative distribution function. $F_{\bar{p}_r}^-$ is the inverse of function $F_{\bar{p}_r}$. The Wasserstein distance used for the model training is the rightmost part of equation \ref{eq:final_ws}. Fig. \ref{fig:ws_work} shows the interpretation of the loss function. To calculate the value of the Wasserstein distance between two signals, we first get the cumulative sum of each signal. From equation \ref{eq:final_ws}, the Wasserstein distance is the absolute difference between the two cumulative sum curves, as the gray area shown in the third column of Fig. \ref{fig:ws_work}.

\subsubsection{Analysis of Peak Misalignment}
\begin{figure}
\centering
\begin{minipage}[b]{0.49\linewidth}
  \centering
  \centerline{\includegraphics[width=\linewidth]{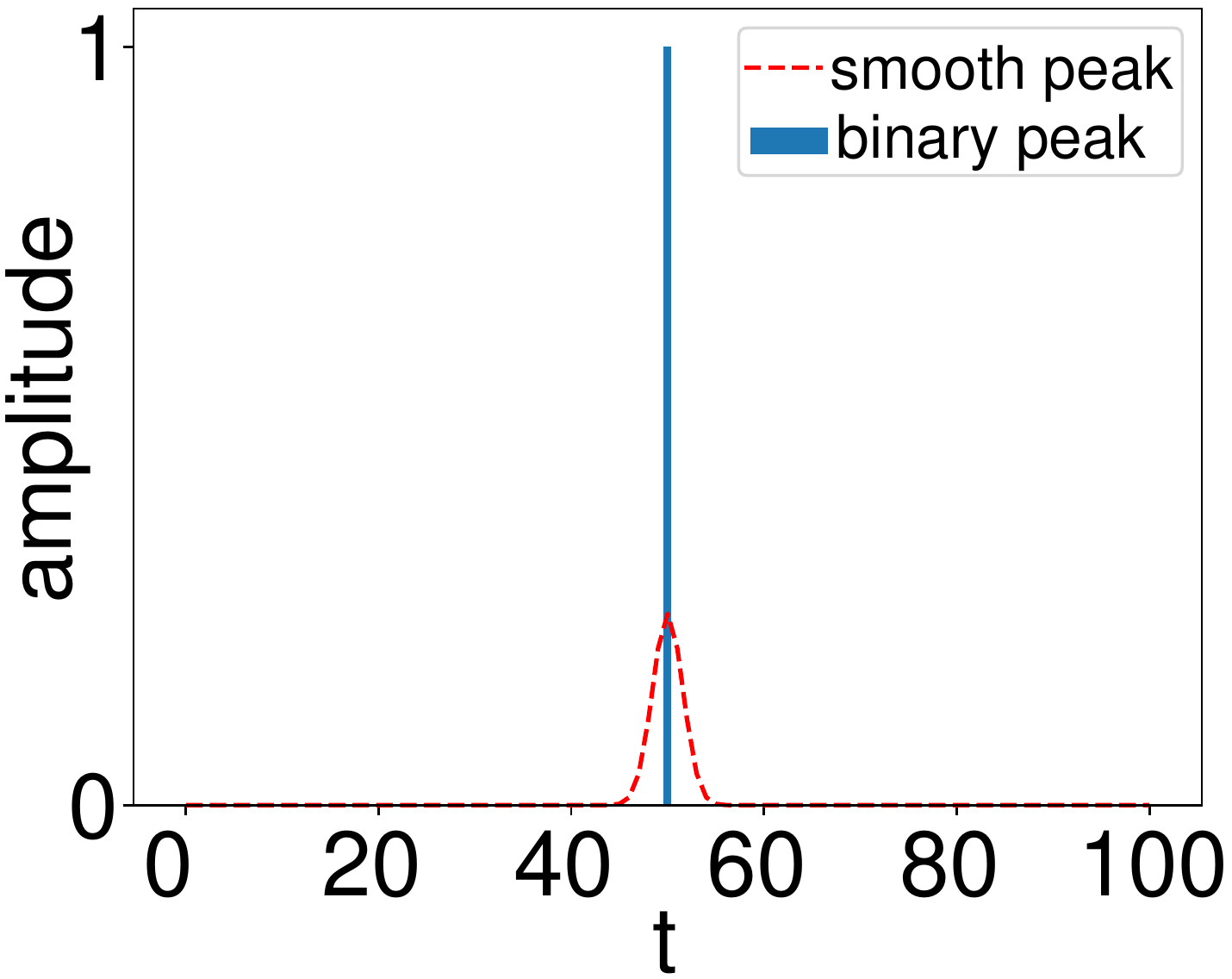}}
  \centerline{(a)}
\end{minipage}
\begin{minipage}[b]{0.49\linewidth}
  \centering
  \centerline{\includegraphics[width=\linewidth]{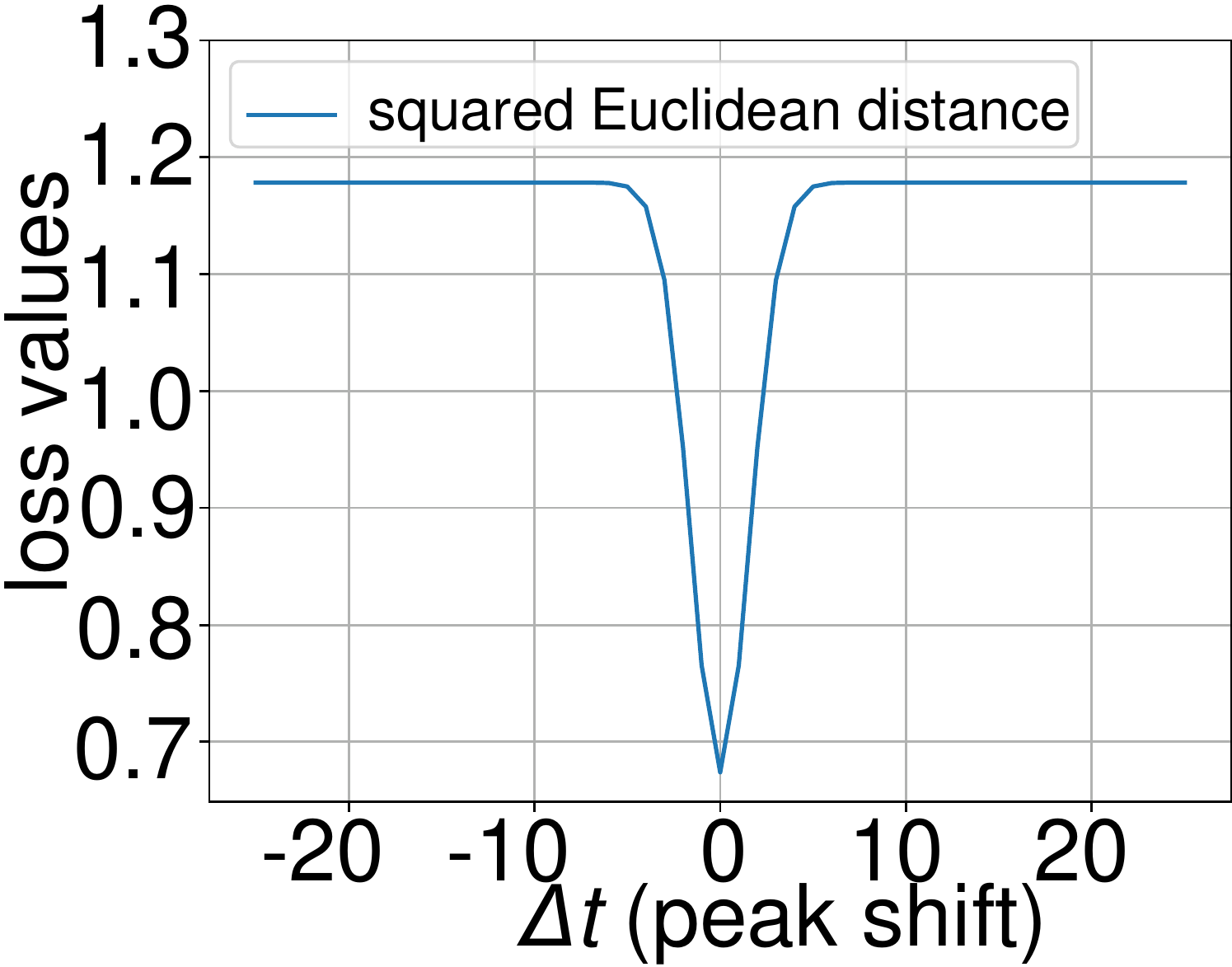}}
  \centerline{(b)}
\end{minipage}

\begin{minipage}[b]{0.49\linewidth}
  \centering
  \centerline{\includegraphics[width=\linewidth]{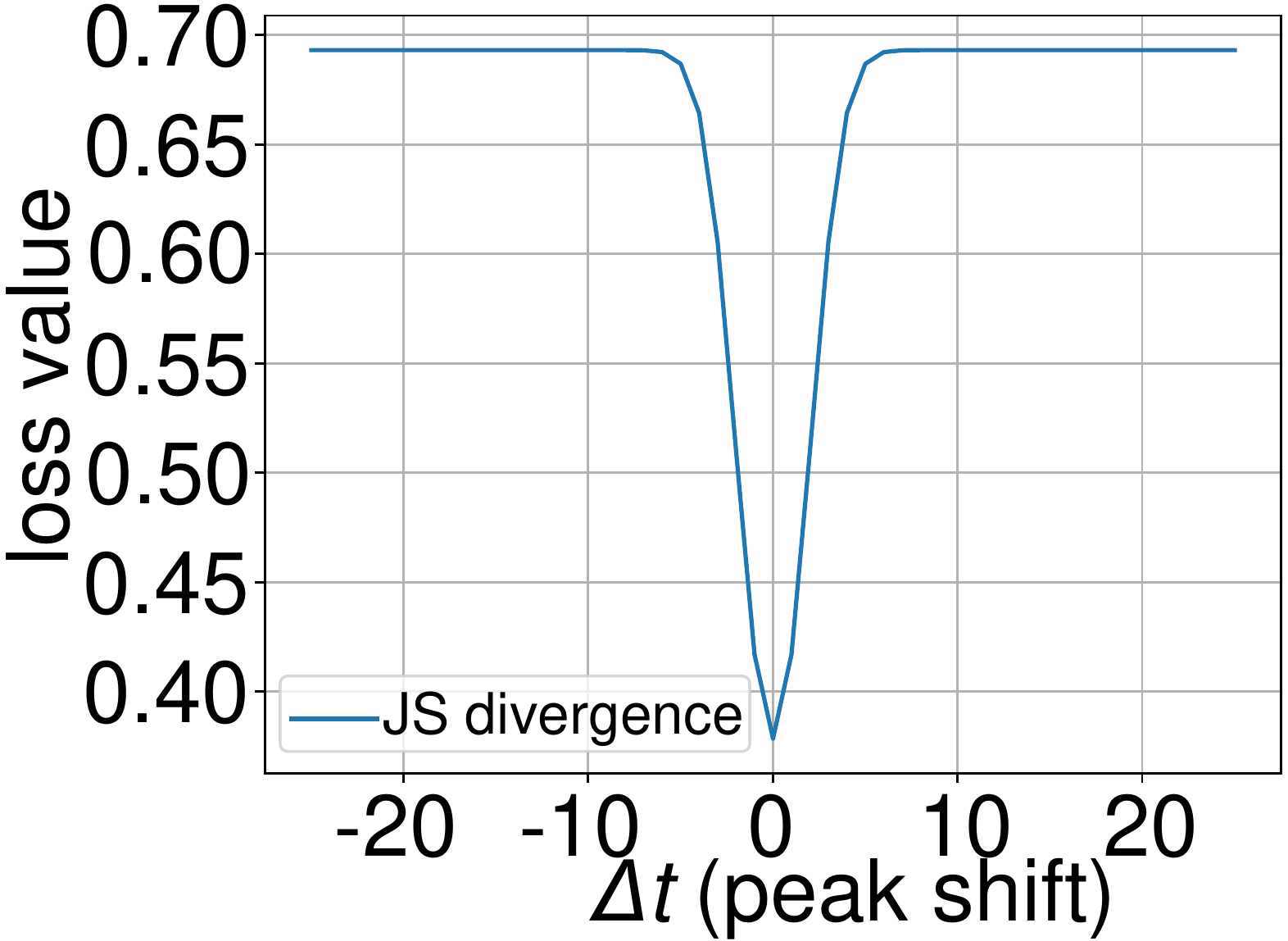}}
  \centerline{(c) }
\end{minipage}
\begin{minipage}[b]{0.49\linewidth}
  \centering
  \centerline{\includegraphics[width=\linewidth]{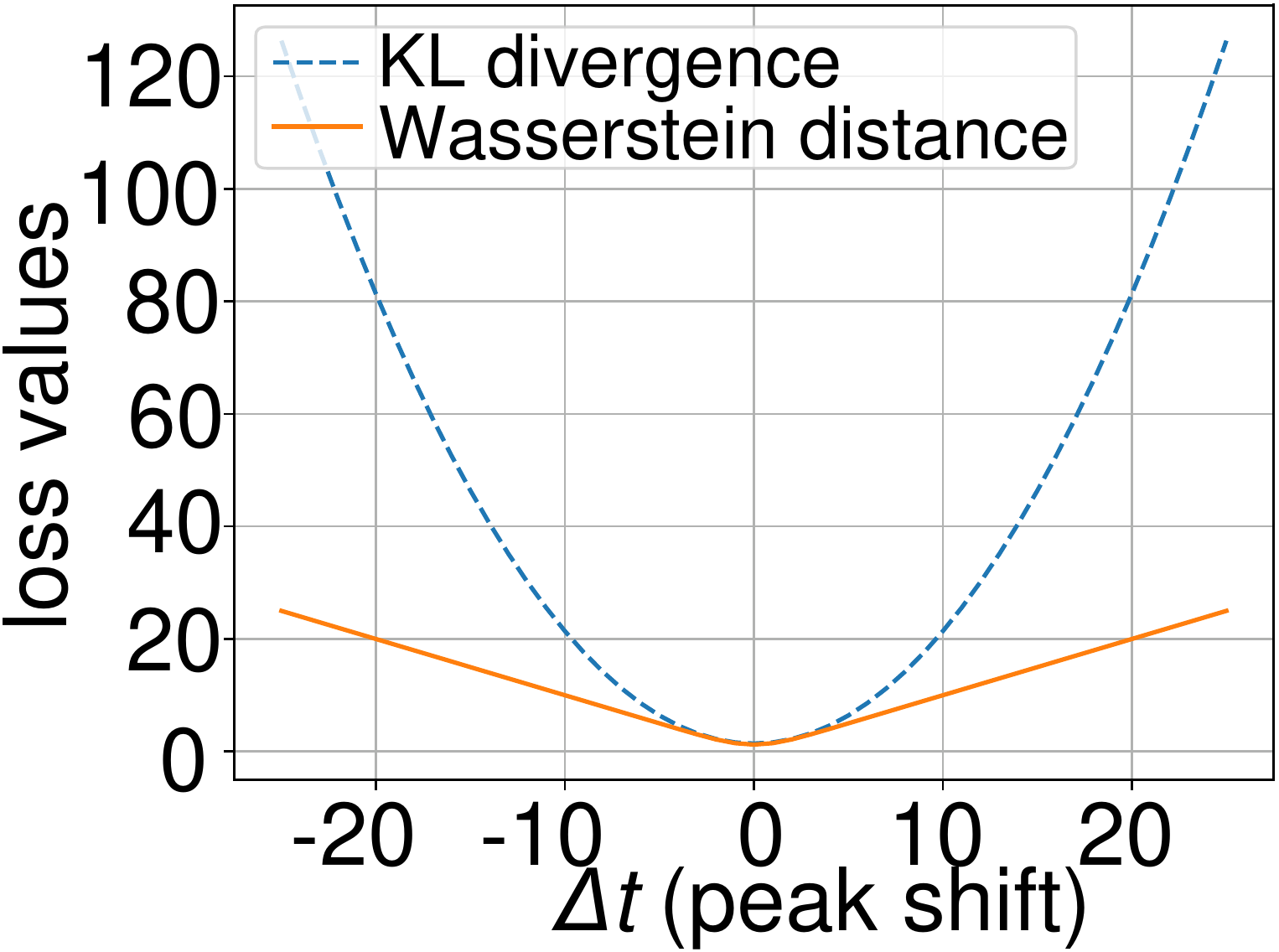}}
  \centerline{(d)}
\end{minipage}

\caption{(a) The smooth peak and binary peak. (b) Squared Euclidean Distance with respect to the peak shift $\Delta t$. (c) JS divergence with respect to the peak shift $\Delta t$. (d) KL divergence and Wasserstein distance with respect to the peak shift $\Delta t$.}
\label{fig:peak_shift}
\end{figure}

We first analyze the loss function response when two comparing signals are misaligned. We first generate a smooth peak signal $p_s$ from a truncated Gaussian distribution curve $\mathcal{N}(0, \sigma^2)$ with $\sigma^2=0.1$ to imitate the 3DCNN model output and normalize it to make sure the sum is one. We also generate a binary peak signal $p_b$ to imitate the ground truth systolic peak. These two signals are shown in Fig. \ref{fig:peak_shift}(a). We shift the binary peak signal $p_b(t)$ to $p_{b, \Delta t}(t)=p_b(t-\Delta t)$ and get the loss function values $l(p_s, p_{b, \Delta t})$. We can plot the loss function values $l(p_s, p_{b, \Delta t})$ with respect to peak shift $\Delta t$ to see the loss functions response to the peak misalignment. The results are shown in Fig. \ref{fig:peak_shift}(b-d). For squared Euclidean distance and JS divergence in Fig. \ref{fig:peak_shift}(b-c), the loss value becomes saturated and constant when the absolute peak shift is too large, which means the loss function cannot assign a larger penalty to a larger peak alignment. This disadvantage might prevent the model from learning the accurate systolic peaks at these saturated locations. On the other side, KL divergence and Wasserstein distance in Fig. \ref{fig:peak_shift}(d) assign a larger penalty when the peak misalignment is larger and will not be saturated when the peak shift is large, which means KL divergence and Wasserstein distance could be the promising options.

\subsubsection{Analysis of Peak Sharpness}
\begin{figure}
\centering
\begin{minipage}[b]{0.49\linewidth}
  \centering
  \centerline{\includegraphics[width=\linewidth]{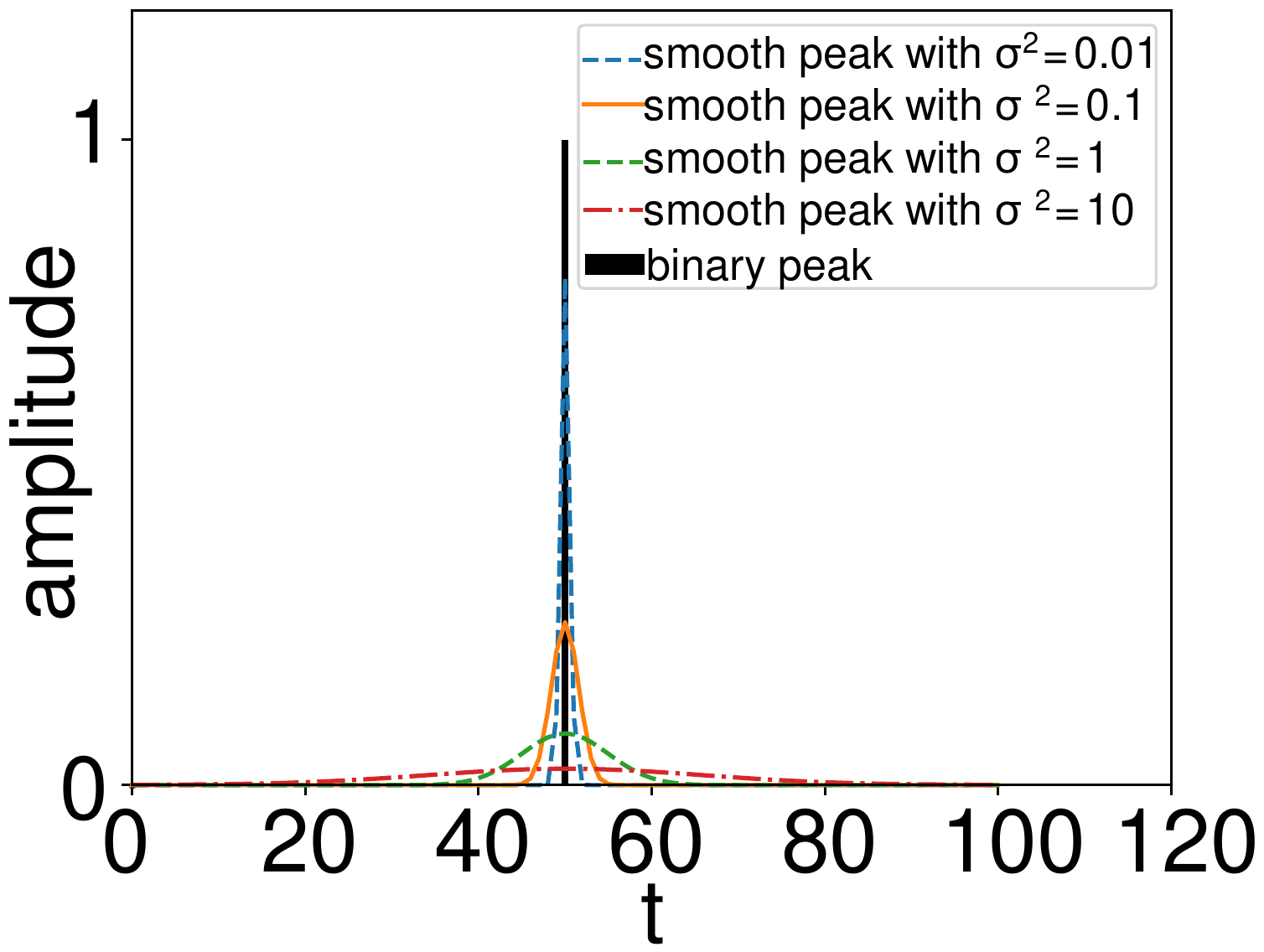}}
  \centerline{(a)}
\end{minipage}
\begin{minipage}[b]{0.49\linewidth}
  \centering
  \centerline{\includegraphics[width=\linewidth]{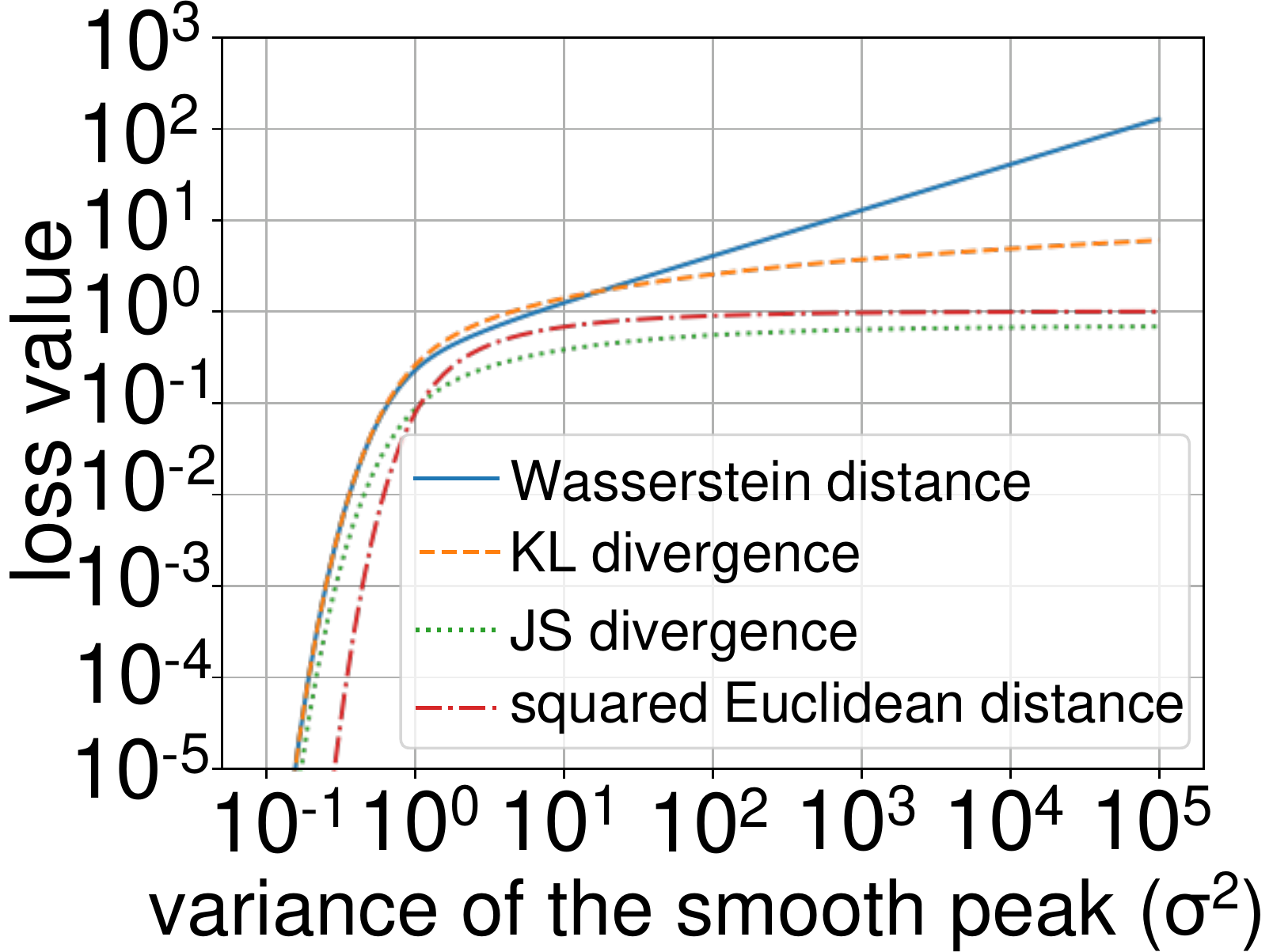}}
  \centerline{(b)}
\end{minipage}

\caption{(a) The binary peak and the smooth peak with difference variances (sharpness). (b) The loss values of the four candidate loss function with respect to the variance $\sigma^2$ of the smooth peak.}
\label{fig:peak_var}
\end{figure}

We also analyze the loss function response to peak sharpness. We hope the 3DCNN model will output signals with sharper peaks rather than flat ones, as sharper peaks are less impacted by noises and easier to detect. We use the binary peak and a series of smooth peak signals with different variances $\sigma^2$ to analyze the response of the loss function to the peak sharpness. We can change the variance $\sigma^2$ of the truncated Gaussian distribution curve $\mathcal{N}(0, \sigma^2)$ to control the sharpness of the smooth peak $p_{s, \sigma^2}$. Smaller variance $\sigma^2$ means sharper peaks. This phenomenon is illustrated in Fig. \ref{fig:peak_var}(a). The smooth peak with $\sigma^2=10$ is almost flat, while the smooth peak with $\sigma^2=0.01$ is very sharp and similar to the binary peak. By changing the variance $\sigma^2$ of the smooth peak, we can plot the loss values $l(p_{s, \sigma^2}, p_{b})$ with respect to the variance $\sigma^2$ of the smooth peak in Fig. \ref{fig:peak_var}(b). From the plot, Wasserstein distance has larger penalty values than other loss functions, especially at large variance positions (small sharpness), which indicates that Wasserstein distance can assign a much larger penalty value when the output peaks are too flat. Therefore, Wasserstein distance is the best to motivate the model to produce sharp peaks.

\subsection{AF detection with HRV features}
\label{sec:hrv}

The AF detection part is shown in the right part of Fig. \ref{fig:framework}. We select 20 features as listed below according to previous studies \cite{lee2012atrial, eerikainen2019detecting, couderc2015detection, li2018obf, shi2019atrial} to train a support vector machine (SVM) classifier with radial basis function (RBF) kernel.

\begin{itemize}
\item Time domain: mean inter-beat interval (IBI), standard deviation of IBI (SDNN), standard deviation of successive difference of IBI (SDSD), percentage of samples with more than 50 ms difference from the consecutive beat (pNN50), percentage of samples with more than 20 ms difference from the consecutive beat (pNN20), the number of samples with more than 50 ms difference from the consecutive beat (NN50), the number of samples with more than 20 ms difference from the consecutive beat (NN20), the root mean square of successive differences of IBI (RMSSD), median of IBI, range of IBI, the coefficient of variation of successive differences (CVSD), the coefficient of variation (CVNNI), maximum heart rate, minimum heart rate, and the standard deviation of heart rate.
\item Spectral domain: the power in the low frequency (LF) (0.04Hz-0.15Hz), the power in the high frequency (HF) (0.15Hz-0.4Hz), and the ratio of LF and HF.
\item Geometrical domain: Poincaré plot standard deviations (SD1, SD2).
\end{itemize}

\section{Experimental Setup}

In this section, we first present our dataset. We will provide the experimental protocol and evaluation metrics for two experiments. One experiment is remote PPG monitoring, and another is non-contact AF/AFL detection.

\subsection{Dataset}

\subsubsection{Participants}
We record the full version of the Oulu Bio-face (OBF) dataset with 100 healthy subjects and 100 AF patients. The healthy subjects were recruited from the University of Oulu, and AF patients were from Oulu University Hospital. The study was performed according to the Declaration of Helsinki, and the local committee of research ethics of the Northern Ostrobothnia Hospital District approved the protocol (reference number: 116/2016). All participants gave their written consent before recording. The statistical information of participants is summarized in Table \ref{tb:participants}. The partitions of healthy subjects and patients in the OBF dataset are OBF-H and OBF-P, respectively.

\begin{table}[h]
\caption{Statistical Information of the participants}
\centering
\begin{tabular}{@{}l|c|c@{}}
\toprule
                & OBF-H (n=100)                                                                        & OBF-P (n=100)            \\ \midrule
Age (year)             & 31.6 ± 8.8, {[}18, 68{]}                                                               & 64.2 ± 9.2, {[}43, 88{]} \\ \midrule
Gender          & \begin{tabular}[c]{@{}l@{}}61\% Male\\39\% Female\end{tabular}                                                                        & \begin{tabular}[c]{@{}l@{}}83\% Male\\17\% Female\end{tabular}             \\ \midrule
Ethnic          & \begin{tabular}[c]{@{}l@{}}Caucasian: 32\%, \\ Asian: 37\%, \\ Others: 31\%.\end{tabular} & Caucasian: 100\%           \\ \midrule
Weight (kg)         & 71 ± 16                                                                                & 96 ± 21                    \\ \midrule
Ratio of wearing eyeglasses & 39\%                                                                                   & 45\%                       \\ \bottomrule
\end{tabular}
\label{tb:participants}
\end{table}

\subsubsection{Data Acquisition}

\begin{figure}[h]
\begin{minipage}[b]{\linewidth}
  \centering
  \centerline{\includegraphics[width=\linewidth]{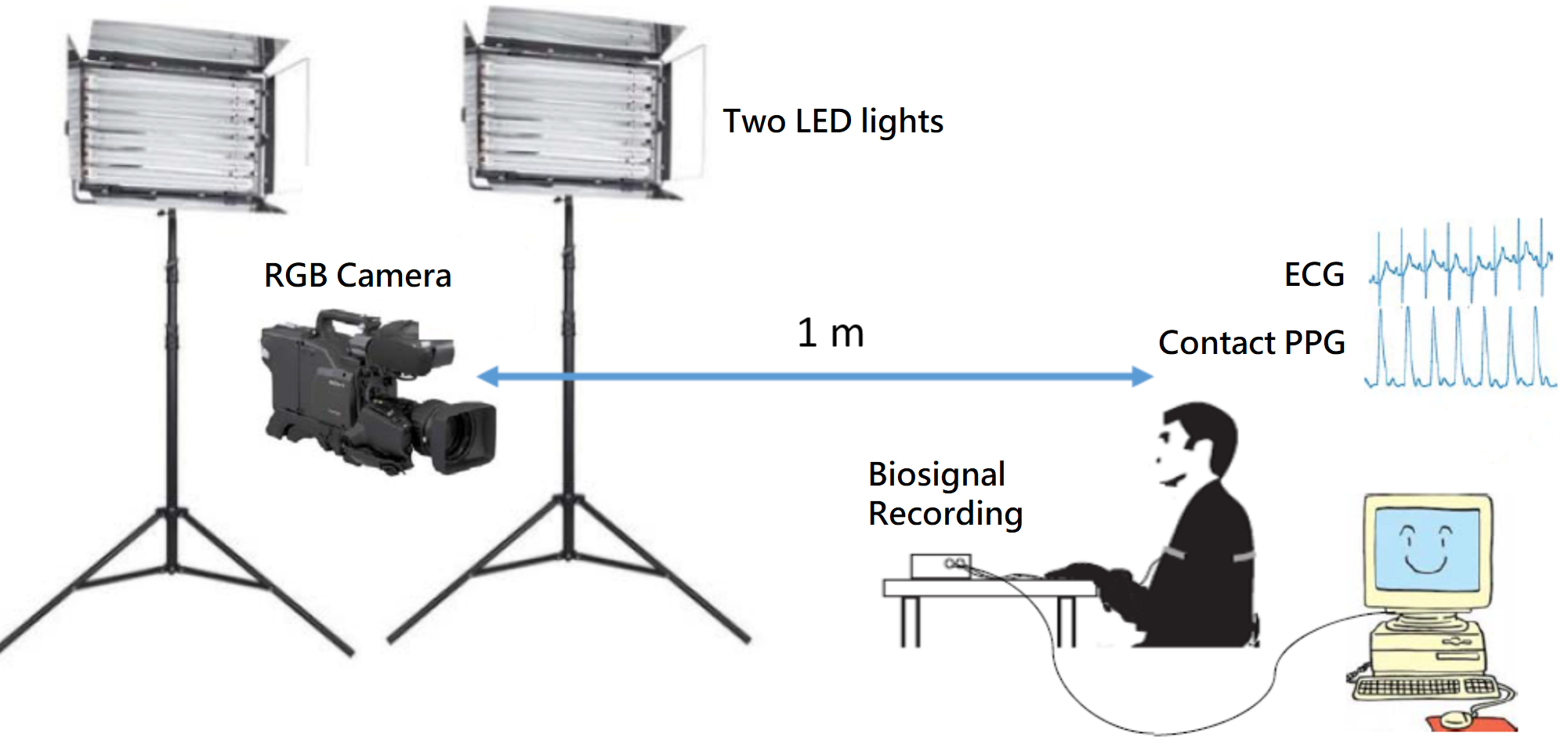}}
\end{minipage}

\caption{Data Recording Setup. RGB = red green blue, ECG = electrocardiography, PPG = photoplethysmography, LED = light-emitting diode}

\label{fig:recording_setup}
\end{figure}

The recording setup is shown in Fig. \ref{fig:recording_setup}. A participant was seated in front of an RGB camera at a one-meter distance. Two LED lights on both sides of the RGB camera face toward the participant at a 45-degree angle at a 1.5-meter distance. Meanwhile, ECG and contact PPG signals are measured by ECG and PPG sensors, respectively, and synchronized with the video. The recording equipment and settings are shown in Tab. \ref{tb:equipment}.

\begin{table}[h]
\caption{Recording Equipment and Settings}
\centering
\begin{threeparttable}
\begin{tabular}{@{}cccl@{}}
\toprule
Devices                                                        & Speicifications                                                      & Settings                                                                                 & Outputs                                                        \\ \midrule
RGB camera                                                    & \begin{tabular}[c]{@{}c@{}}Blackmagic \\ URFA mini\end{tabular}      & \begin{tabular}[c]{@{}c@{}}ISO 400, \\ FPS 60, \\ Resolution \\ 1920 x 1080\end{tabular} & RGB video                                                     \\ \midrule
ECG sensor                                                    & NX-EXG2B                                                             & \begin{tabular}[c]{@{}c@{}}Sampling rate \\ 256 Hz\end{tabular}                          & ECG signal                                                    \\ \midrule
\begin{tabular}[c]{@{}c@{}}Contact \\ PPG sensor\end{tabular} & NX-BVP1C                                                             & \begin{tabular}[c]{@{}c@{}}Sampling rate \\ 128Hz\end{tabular}                           & \begin{tabular}[c]{@{}l@{}}Contact \\ PPG signal\end{tabular} \\ \midrule
LED lights                                                    & \begin{tabular}[c]{@{}c@{}}Aputure, \\ LightStorm LS 1c\end{tabular} & \begin{tabular}[c]{@{}c@{}}Brightness: 3500 lux, \\ Temperature: 5500 k\end{tabular}     & N/A                                                           \\ \midrule
                                                              &                                                                      &                                                                                          &                                                               \\
                                                              &                                                                      &                                                                                          &                                                              
\end{tabular}
\vspace{-0.8cm}
\begin{tablenotes}
\item RGB = red green blue, ECG = electrocardiography, PPG = photoplethysmography, LED = light-emitting diode
\end{tablenotes}
\end{threeparttable}
\label{tb:equipment}
\end{table}

During the recording, the participant was seated facing the camera, and the ECG/contact PPG sensors were attached to them. For each healthy subject, there are two recording sessions, and each lasts for five minutes. One session was recorded at a resting status, and the other was recorded after 5 minutes of exercise (climbing the stairs) so that a wider range of heart rates is covered. The mean heart rates before and after exercise are 73.6 bpm and 83.2 bpm, respectively. The heart rate ranges before and after exercise are 50.09-107.2 bpm and 60.64-130.1 bpm, respectively. For each AF patient, there are also two recording sessions. One was recorded before cardioversion treatment (AF symptom presented), and the other was recorded after the treatment (back to sinus rhythm) so that each patient has both AF and SR data. The mean heart rates before and after treatment are 78.9 bpm and 66.5 bpm, respectively. The corresponding ECG and contact PPG were recorded simultaneously with each video. Finally, OBF-H has 200 videos. Due to data loss, some patients only have one session, and the total number of videos in OBF-P is 169.

\subsubsection{Data Labelling}

The dataset composition and labels are shown in Table \ref{tb:obf_dataset_structure}. In order to do the AF detection, all the videos in OBF-H are labeled as healthy. All patients' data in OBF-P were labeled by two independent cardiologists by observing ECG signals. Only persistent AF/SR/AFL (i.e., no rhythm transition during the recording) cases were labelled as AF/SR/AFL, and complex cases (e.g., paroxysmal, ventricular extrasystole, supraventricular extrasystoles) were labelled as 'Other' (24 cases). Finally, there are 73 videos labeled as AF, 61 videos labeled as SR, 11 videos labeled as AFL, and 24 videos labeled as other that are not used for classification experiments.




\begin{table}[]
\caption{OBF Dataset Composition and Classification Labels\\(The numbers of participants are in the brackets)}
\centering
\begin{threeparttable}
\begin{tabular}{|c|c|c|c|c|}
\hline
\bf Modalities& \begin{tabular}[c]{@{}c@{}}\bf Dataset\\\bf Partitions\end{tabular}     & \bf Participants & \bf Sessions & \bf Labels \\ \hline
\multirow{4}{*}{\begin{tabular}[c]{@{}c@{}}RGB Videos,\\ ECG,\\ Contact PPG\end{tabular}} & \multirow{2}{*}{OBF-H} & \multirow{2}{*}{\begin{tabular}[c]{@{}c@{}} Healthy\\Subjects\end{tabular}} & \begin{tabular}[c]{@{}c@{}}Before\\ Exercise\\ (100)\end{tabular}     & \multirow{2}{*}{\begin{tabular}[c]{@{}c@{}}Healthy\\ (200)\end{tabular}} \\ \cline{4-4}
 & & &\begin{tabular}[c]{@{}c@{}}After\\ Exercise\\ (100)\end{tabular}      &                                                                                                     \\ \cline{2-5} 
 & \multirow{2}{*}{OBF-P} & \multirow{2}{*}{\begin{tabular}[c]{@{}c@{}}AF\\Patients\end{tabular}} & \begin{tabular}[c]{@{}c@{}}Before\\ Cardioversion\\ (90)\end{tabular} & \multirow{2}{*}{\begin{tabular}[c]{@{}c@{}}AF (73)\\ SR (61)\\ AFL (11)\\ Other (24)\end{tabular}} \\ \cline{4-4}
 & & & \begin{tabular}[c]{@{}c@{}}After\\ Cardioversion\\ (79)\end{tabular}  &                                                                                                     \\ \hline
\end{tabular}
\begin{tablenotes}
\item RGB = red green blue, ECG = electrocardiography, PPG = photoplethysmography, AF = atrial fibrillation, SR = sinus rhythm, AFL = atrial flutter. 
\end{tablenotes}
\end{threeparttable}
\label{tb:obf_dataset_structure}
\end{table}

\subsection{Experiment I: Remote PPG Monitoring}
\subsubsection{Experimental Protocol}
We train our models with three dataset partitions: OBF-P, OBF-H, or the whole OBF dataset. We also test our models on these three dataset partitions. The model is trained for 45 epochs with a learning rate of 0.0001. Each training video clip has 512 frames, and the batch size for training is 4. The binary systolic peaks are obtained by NeuroKit2 \cite{makowski2021neurokit2} from our contact PPG and will be used as the ground truth to train the 3DCNN model. Wasserstein distance is used as the loss function. For testing, the videos are divided into non-overlapping 30-seconds clips and used as inputs. We use subject independent 10-folds cross-validation in all experiments except two cases, i.e., training on OBF-H and testing on OBF-P, and training on OBF-P and testing on OBF-H, which use cross-set protocol. We also compare our method (3DCNN-PEAK) with four remote PPG methods: 3DCNN-BVP \cite{yu2019remote} is a deep learning-based method using contact PPG waveforms for training. POS \cite{wang2016algorithmic}, CHROM \cite{de2013robust}, and PBV \cite{de2014improved} are classical remote PPG algorithms without training.

\subsubsection{Evaluation Metrics}
We evaluate accuracy on two levels, i.e., the average heart rate level and IBI level. Accurate measure on the IBI level is more challenging and essential for AF detection since it requires accurate systolic peaks.  

We use mean absolute error (MAE), root mean squared error (RMSE), and Pearson correlation (R) to evaluate the error of the average heart rate. MAE for heart rate is defined as $\text{MAE}_{HR} = \sum_{n=1}^N |\text{HR}_{video}-\text{HR}_{true}|/N$, where N is the number of samples, $\text{HR}_{video}$ is the heart rate measured from face videos, and $\text{HR}_{true}$ is the true heart rate obtained from the contact PPG. RMSE is defined as $\text{RMSE}_{HR} = \sqrt{\sum_{n=1}^{N} (\text{HR}_{video}-\text{HR}_{true})^2/N} $. Small MAE and RMSE values indicate accurate heart rate estimation. Pearson correlation is the linear correlation between the heart rates measured from videos and the true heart rates. When the Pearson correlation is close to 1, the heart rate estimation is accurate. 

We use three metrics of mean absolute error (MAE), standard deviation (STD), and accuracy as used in \cite{liu2020detecting} to evaluate the error of IBI. We can first define the absolute error of IBI as $\text{AE} = \sum_{t=1}^K|\text{IBI}_{video}(k)-\text{IBI}_{true}(k)|/K$ where K is the length of the IBI, $\text{IBI}_{video}$ is the IBI curve from face video, and $\text{IBI}_{true}$ is the true IBI curve from contact PPG. Since the original IBI curve is an irregularly spaced time series, we should resample the original IBI curve to get an evenly spaced time series to calculate the absolute error between two IBI curves. The MAE and STD of IBI are the mean and standard deviation of $\text{AE}$ for all samples. The accuracy of IBI is defined as $\text{AC}_{IBI}=1- \frac{1}{N}\sum_{n=1}^N \text{AE}_n/(T/(B_n-1))$ where $T$ is the time length of a IBI curve and $B_n$ is the number of systolic peaks in the $n^{th}$ sample.

\subsection{Experiment II: AF/AFL detection}
\subsubsection{Experimental Protocol}
Remote PPG is obtained from the model trained on both OBF-P and OBF-H. HRV features (as described in Section \ref{sec:hrv}) are calculated from the systolic peaks in remote PPG and used for AF detection experiments. We perform two kinds of AF detection. 1) We use videos from healthy subjects and videos with AF labels from patients to classify healthy vs. AF. 2) We use the patient videos with SR labels and patient videos with AF labels to classify SR vs. AF. We perform the classification with different clip lengths of 10s, 20s, 30s, 60s, and 120s to see how the clip length influences the AF detection. 3) We also perform the classification of SR vs. AFL with clip length 30s to investigate whether AFL can be detected from face videos.

In the classification experiments of healthy vs. AF and SR vs. AF, we use subject-independent 10-fold cross-validation \footnote{This means that all subjects are evenly divided into 10 folds. All data from one subject is only in one fold. For each testing, one fold is used for testing, and the left nine folds are used for training, which means data from one subject is never in both testing and training.}. In the classification experiment of SR vs. AFL, the numbers of two labels are unbalanced, and the number of samples is limited, so we use another evaluation protocol in \cite{shi2019atrial}. We randomly select 6 SR patients and 6 AFL patients to form the training set while randomly selecting another 5 SR patients and the left 5 AFL patients as the testing set. Thus, subjects in the training and testing sets are balanced and not overlapped. The experiments are independently performed ten times, and the average performance is reported.

We also report results achieved by using the ECG signals and contact PPG signals with the same HRV features in Sec. \ref{sec:hrv}, which are the ECG/contact PPG reference methods. Theoretically, the ECG and contact PPG results should be the upper bound that the model can achieve and are the reference results compared with remote PPG results. We also report results of ECG-based AF detection methods \cite{islam2019robust,fan2018multiscaled} for healthy vs. AF and SR vs. AF. For remote PPG, we compare our results with four previous methods\cite{yu2019remote,shi2019atrial, wang2016algorithmic, de2014improved}. For the baseline methods \cite{yu2019remote, wang2016algorithmic, de2014improved} and ECG/contact PPG reference methods, we use the same SVM classifier and HRV features from Sec. \ref{sec:hrv}, while only change the input IBI series. For \cite{islam2019robust,fan2018multiscaled,shi2019atrial}, we use the features and models in their papers.

\subsubsection{Evaluation Metrics}
AF detection results are summarized into true positive (TP), true negative (TN), false positive (FP), and false negative (FN). We use accuracy, sensitivity, and specificity as the classification metrics. Accuracy is defined as $(TP+TN)/(TP+TN+FP+FN)$. Sensitivity is defined as $TP/(TP+FN)$. Specificity is defined as $TN/(TN+FP)$. We also add metrics of F1 score and area under curve (AUC) for more convincing comparison.

\section{Results and Discussion}

In this section, we will present the results of our two experiments. One experiment is remote PPG monitoring, and another is non-contact AF/AFL detection. Results about computational speed are also reported. Finally, extensive discussion about the results is provided.

\begin{figure*}[h]
\centering
\begin{minipage}[b]{0.32\linewidth}
  \centering
  \centerline{\includegraphics[width=\linewidth]{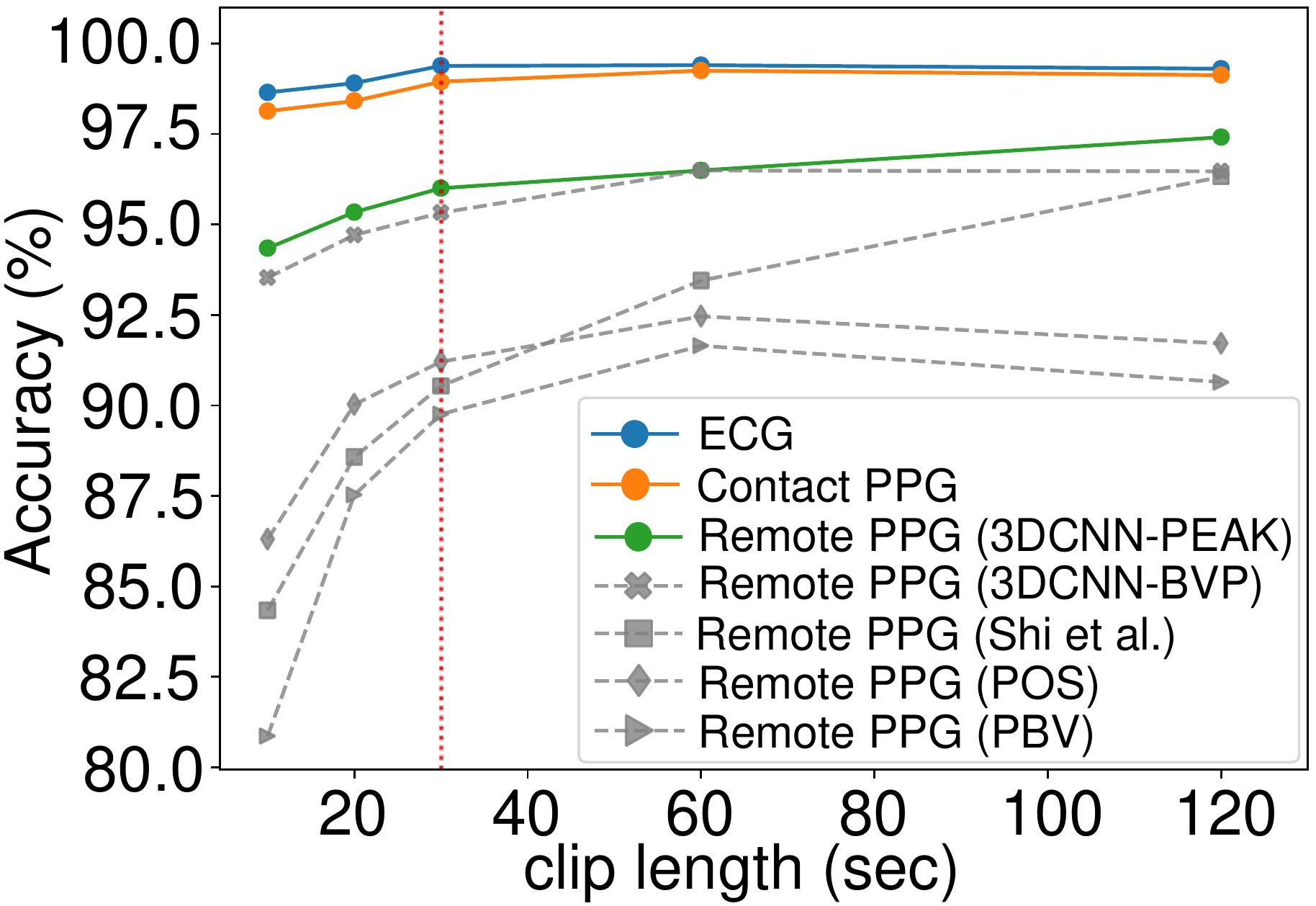}}
  \centerline{(a)}
\end{minipage}
\begin{minipage}[b]{0.32\linewidth}
  \centering
  \centerline{\includegraphics[width=\linewidth]{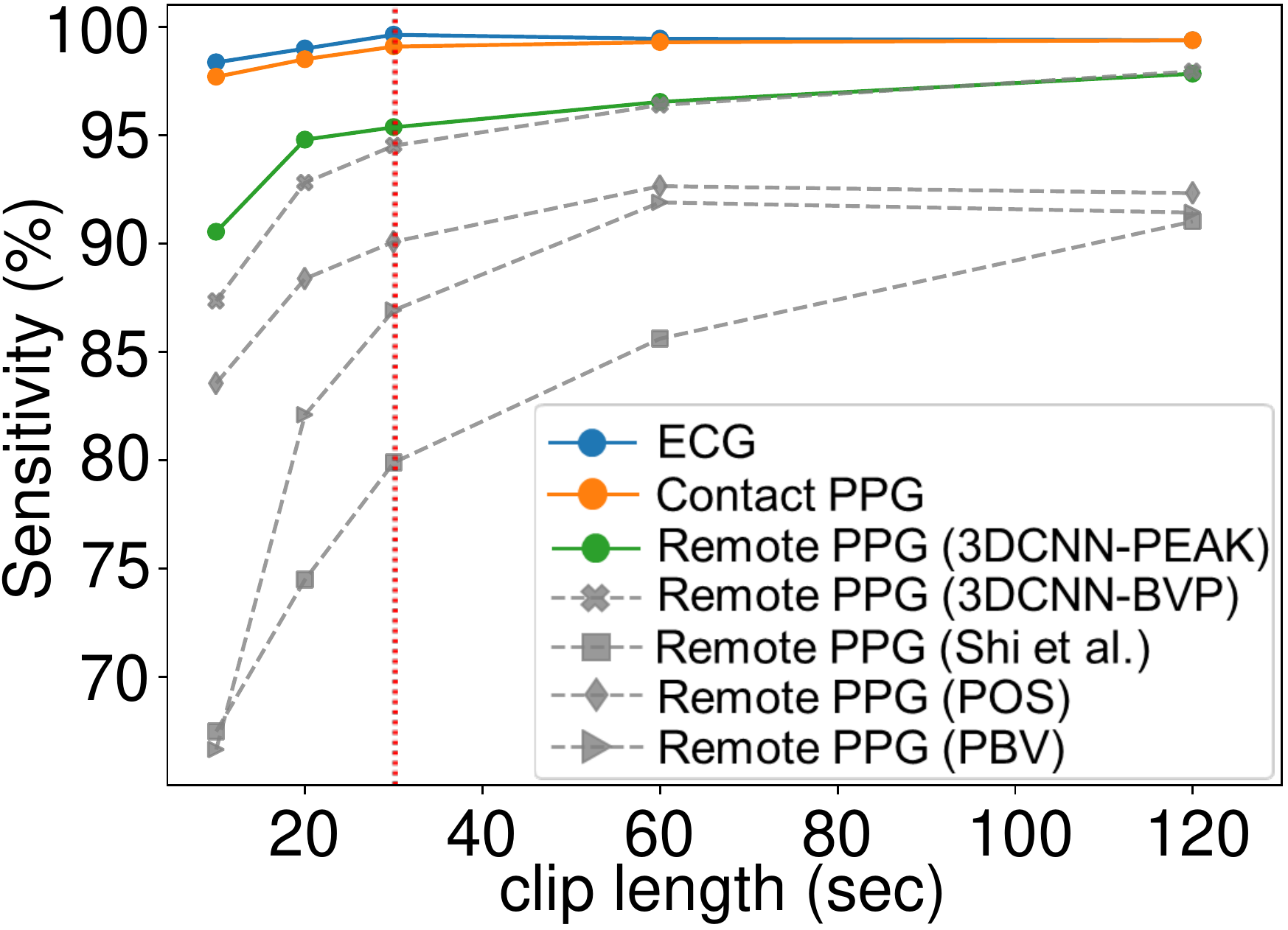}}
  \centerline{(b)}
\end{minipage}
\begin{minipage}[b]{0.32\linewidth}
  \centering
  \centerline{\includegraphics[width=\linewidth]{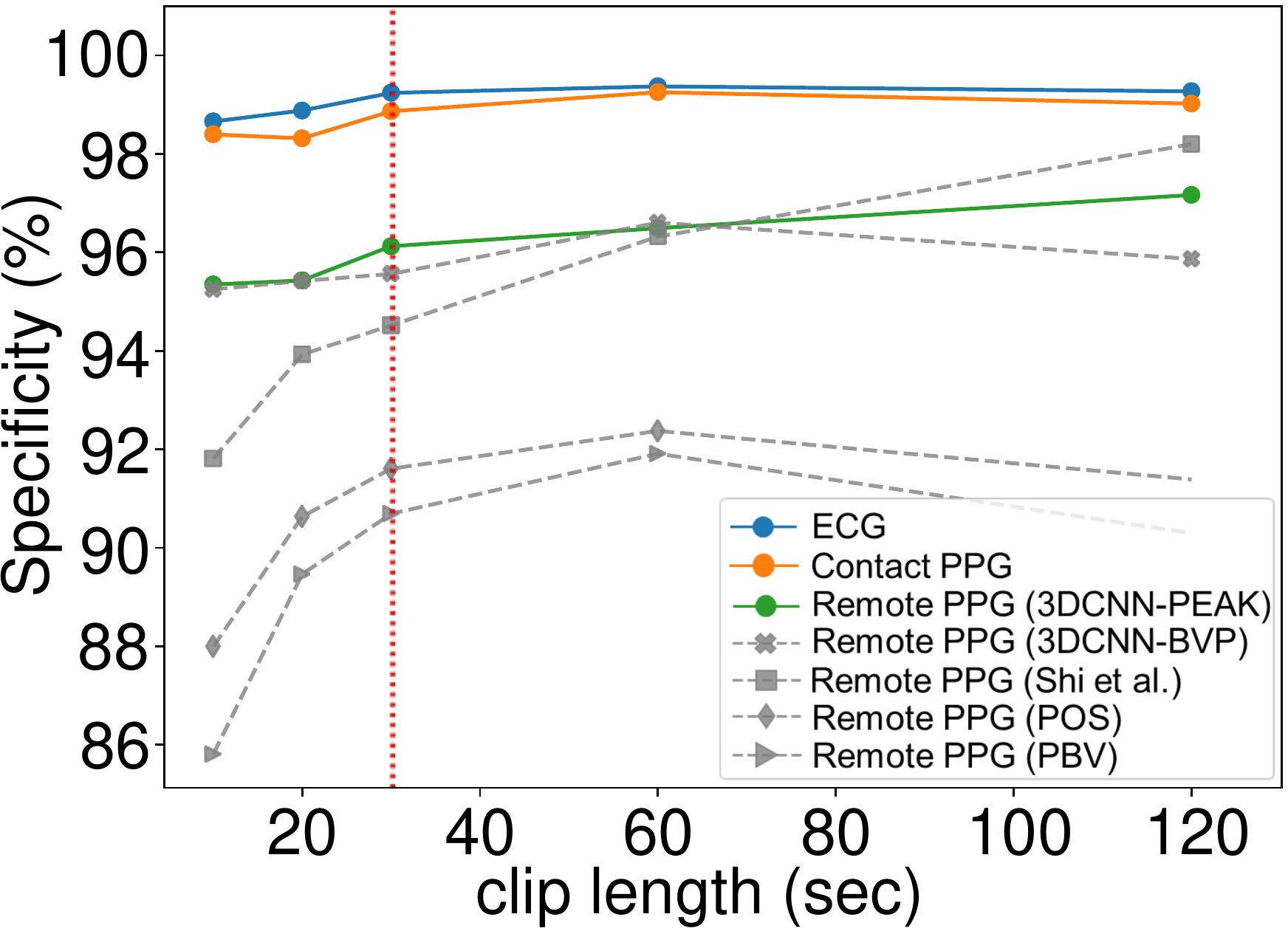}}
  \centerline{(c)}
  \end{minipage}
  
  \begin{minipage}[b]{0.32\linewidth}
  \centering
  \centerline{\includegraphics[width=\linewidth]{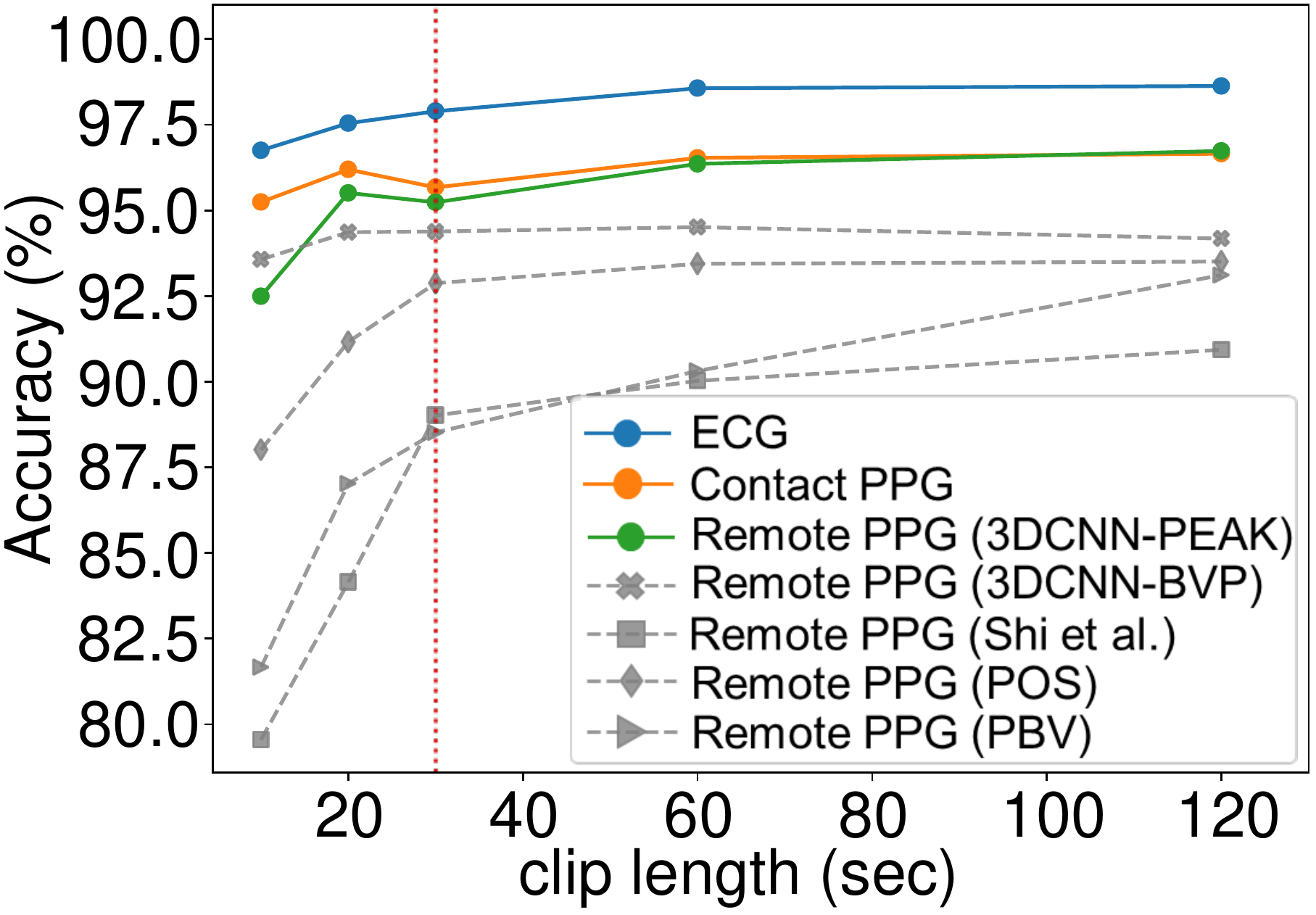}}
  \centerline{(d)}
\end{minipage}
\begin{minipage}[b]{0.32\linewidth}
  \centering
  \centerline{\includegraphics[width=\linewidth]{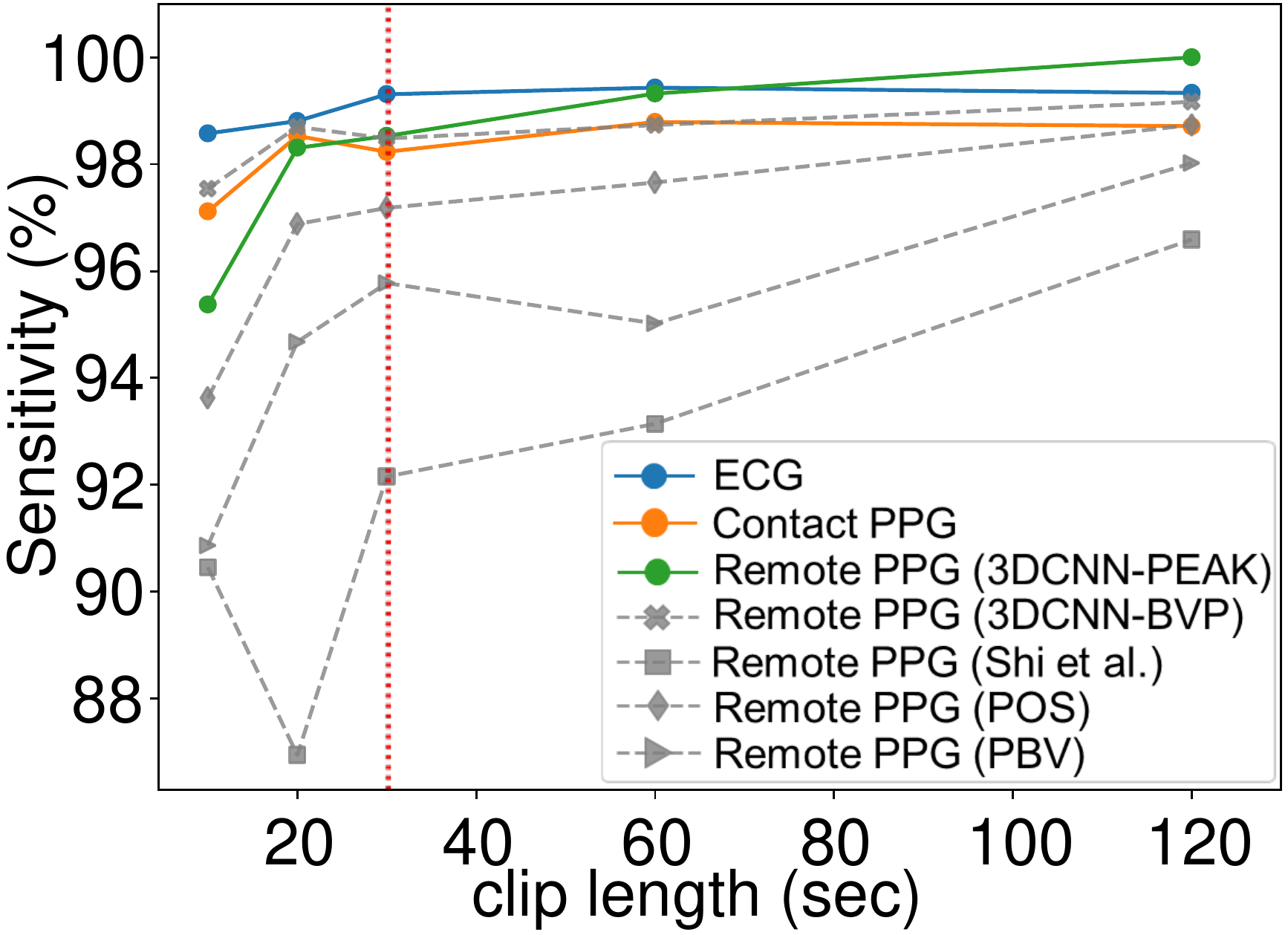}}
  \centerline{(e)}
\end{minipage}
\begin{minipage}[b]{0.32\linewidth}
  \centering
  \centerline{\includegraphics[width=\linewidth]{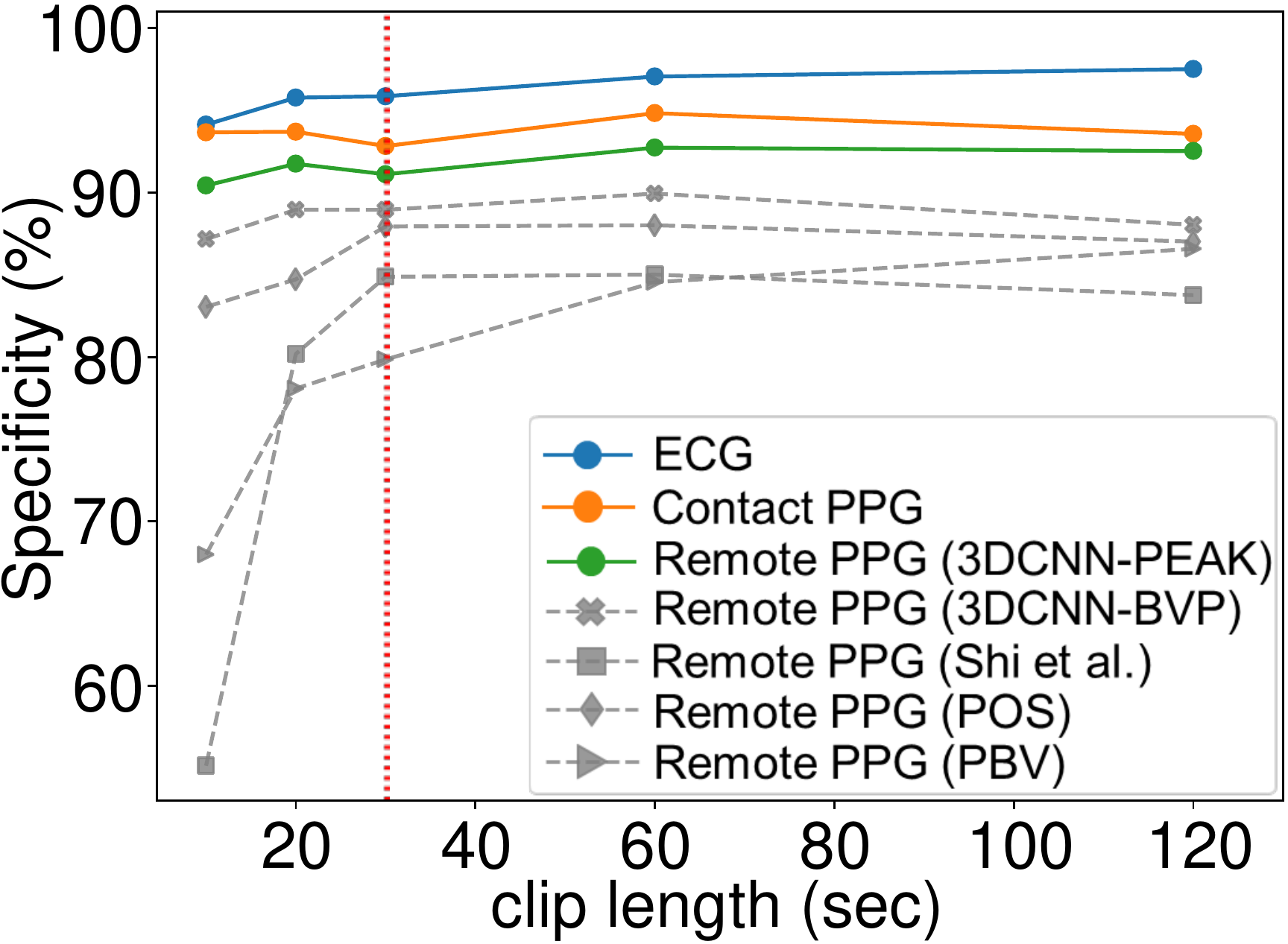}}
  \centerline{(f)}
\end{minipage}

\caption{(a-c) The accuracy, sensitivity, and specificity with respect to the clip length for healthy vs. AF. (d-f) The accuracy, sensitivity, and specificity with respect to the clip length for SR vs. AF. The red dot lines are the 30s clip length that we choose for our classification experiments. ECG = electrocardiography, PPG = photoplethysmography.}
\label{fig:AF_clip}
\end{figure*}

\subsection{Results for Experiment I: Remote PPG Monitoring}

We compare these four candidate loss functions mentioned in Sec. \ref{sec:loss}. Table \ref{tb:comparision_loss} shows that Wasserstein distance has the best performance for remote PPG monitoring. Fig. \ref{fig:ws_work} also shows how Wasserstein distance boosts the predicted systolic peaks. Therefore, our experimental results validate our analysis conclusion in Section \ref{sec:loss_select} that Wasserstein distance is the best option. 

\begin{table}[h]
\caption{Performance Comparison of the candidate loss functions}
\centering
\begin{threeparttable}
\begin{tabular}{l|c c c | c c c}
\toprule
\multirow{2}{*}{} & \multicolumn{3}{c|}{Heart Rate}                                                                                & \multicolumn{3}{c}{Inter-beat Interval}                                                                     \\ \cmidrule{2-7} 
                  & \begin{tabular}[c]{@{}c@{}}MAE\\ (bpm)\end{tabular} & \begin{tabular}[c]{@{}c@{}}RMSE\\ (bpm)\end{tabular} & R      & \begin{tabular}[c]{@{}c@{}}MAE\\ (ms)\end{tabular} & \begin{tabular}[c]{@{}c@{}}STD\\ (ms)\end{tabular} & \begin{tabular}[c]{@{}c@{}}Accuracy\\ (\%)\end{tabular} \\ \midrule
KL      &  8.77 &	16.45 &	0.51	& 189.44	& 435.86	& 83.84 \\ \midrule
JS        & 14.80	& 23.59	& 0.30 & 390.62	& 705.08	& 75.83 \\ \midrule
SED    &  7.18	& 14.39	& 0.58	& 159.77	& 437.77	& 86.07  \\ \midrule
WS     & \textbf{1.46}	& \textbf{3.53}	& \textbf{0.96}	& \textbf{50.74}	& \textbf{68.06}	& \textbf{94.18}  \\ \bottomrule
\end{tabular}
\begin{tablenotes}
\item KL = Kullback-Leibler Divergence, JS = Jensen–Shannon Divergence, SED = Squared Euclidean Distance, WS = Wasserstein Distance, MAE = mean absolute error, RMSE = root mean squared error, R = Pearson correlation, STD = standard deviation.
\end{tablenotes}
\end{threeparttable}
\label{tb:comparision_loss}
\end{table}

Table \ref{tb:test_OBF_P} shows the test results on OBF-P. No matter what training set is used, our proposed method (3DCNN-PEAK) achieves the best performance on heart rate and IBI measurement. Using both OBF-P and OBF-H for training is preferable since this can further improve the model performance compared with using OBF-P or OBF-H alone. The cross dataset test shown in the OBF-H row indicates that our 3DCNN-PEAK method has better generalization capability than the 3DCNN-BVP method. Table \ref{tb:test_OBF_H} shows the test results on OBF-H. Our 3DCNN-PEAK method still outperforms other baseline methods. Using both OBF-P and OBF-H for training also achieves the best performance on OBF-H. The reason is that a larger dataset (OBF-P + OBF-H) can include more face videos with different noise types and facial skin types from both healthy subjects and patients, which can train the model to be more robust to noise and better capture subtle remote PPG. It is obvious that the error of heart rate and IBI on OBF-H in Table \ref{tb:test_OBF_H} is much lower than on OBF-P in Table \ref{tb:test_OBF_P} for all methods, which means the remote PPG monitoring for patients is more difficult than for healthy subjects. We provide the final results of training and testing on the whole OBF dataset in Table \ref{tb:test_all}, and our proposed 3DCNN-PEAK achieves the best results where MAE values of heart rate and IBI are 1.4658 BPM and 50.7481 ms, respectively. 

\begin{table*}[h]
\caption{results of heart rate and inter-beat interval on OBF-P}
\centering
\begin{threeparttable}
\begin{tabular}{l|l|c c c|c c c}
\toprule
\multirow{2}{*}{\begin{tabular}[l]{@{}l@{}}Training Set \\ (Validation Protocol)\end{tabular}}    & \multirow{2}{*}{Remote PPG Algorithm} & \multicolumn{3}{c|}{Heart Rate} & \multicolumn{3}{c}{Inter-beat Interval}\\ \cmidrule{3-8}
&                                       & \begin{tabular}[c]{@{}c@{}}MAE\\ (bpm)\end{tabular} & \begin{tabular}[c]{@{}c@{}}RMSE\\ (bpm)\end{tabular} & R      & \begin{tabular}[c]{@{}c@{}}MAE\\ (ms)\end{tabular} & \begin{tabular}[c]{@{}c@{}}STD\\ (ms)\end{tabular} & \begin{tabular}[c]{@{}c@{}}Accuracy\\(\%)\end{tabular}\\ \midrule
\multirow{2}{*}{\begin{tabular}[l]{@{}l@{}}OBF-P and OBF-H \\ (10-folds)\end{tabular}} & 3DCNN-PEAK                            &\textbf{2.5304}                                              & \textbf{5.0275}                                               & \textbf{0.9409} & \textbf{81.2244}                                            & 88.6318                                            & \textbf{90.91}   \\
                                 & 3DCNN-BVP \cite{yu2019remote}                             & 3.7356                                              & 7.1297                                               & 0.8795 & 95.8779                                            & 122.6897                                           & 89.45    \\ \midrule
\multirow{2}{*}{\begin{tabular}[l]{@{}l@{}}OBF-P \\ (10-folds)\end{tabular}}           & 3DCNN-PEAK                            & 3.3825                                              & 6.7187                                               & 0.8916 & 92.8611                                            & 99.0031                                            & 89.32    \\
                                 & 3DCNN-BVP \cite{yu2019remote}                             & 4.4277                                              & 8.3049                                               & 0.8350 & 104.1833                                           & 107.9021                                           & 88.21    \\ \midrule
\multirow{2}{*}{\begin{tabular}[l]{@{}l@{}}OBF-H \\ (cross dataset)\end{tabular}}           & 3DCNN-PEAK                            & 3.6671                                              & 6.6110                                               & 0.8957 & 91.9969                                            & \textbf{84.8351}                                            & 89.08    \\
                                 & 3DCNN-BVP \cite{yu2019remote}                            & 5.4465                                              & 9.9317                                               & 0.7562 & 109.1921                                           & 107.8691                                           & 86.68    \\  \midrule
\multirow{3}{*}{-}               & POS \cite{wang2016algorithmic}                                  & 6.2537                                              & 9.7026                                               & 0.7836 & 121.1604                                           & 93.5224                                            & 85.99    \\ \cmidrule{2-8}
                                 & CHROM \cite{de2013robust}                                & 7.2949                                              & 10.7810                                              & 0.7223 & 152.2947                                           & 90.2355                                            & 82.40    \\ \cmidrule{2-8}
                                 & PBV  \cite{de2014improved}                                 & 9.2961                                              & 13.0999                                              & 0.6294 & 159.7383                                           & 115.8414                                           & 82.28    \\ \bottomrule
\end{tabular}
\begin{tablenotes}
\item MAE = mean absolute error, RMSE = root mean squared error, R = Pearson Correlation, STD = standard deviation.
\end{tablenotes}
\end{threeparttable}
\label{tb:test_OBF_P}
\end{table*}

\begin{table*}[h]
\caption{results of heart rate and inter-beat interval on OBF-H}
\centering
\begin{threeparttable}
\begin{tabular}{l|l|c c c|c c c}
\toprule
\multirow{2}{*}{\begin{tabular}[l]{@{}l@{}}Training Set \\ (Validation Protocol)\end{tabular}}    & \multirow{2}{*}{Remote PPG Algorithm} & \multicolumn{3}{c|}{Heart Rate}                                                                                & \multicolumn{3}{c}{Inter-beat Interval}                                                                     \\ \cmidrule{3-8}
                                 &                                       & \begin{tabular}[c]{@{}c@{}}MAE\\ (bpm)\end{tabular} & \begin{tabular}[c]{@{}c@{}}RMSE\\ (bpm)\end{tabular} & R      & \begin{tabular}[c]{@{}c@{}}MAE\\ (ms)\end{tabular} & \begin{tabular}[c]{@{}c@{}}STD\\ (ms)\end{tabular} & \begin{tabular}[c]{@{}c@{}}Accuracy\\(\%)\end{tabular} \\ \midrule
\multirow{2}{*}{\begin{tabular}[l]{@{}l@{}}OBF-P and OBF-H \\ (10-folds)\end{tabular}} & 3DCNN-PEAK                            & \textbf{0.4534}                                              & \textbf{1.2868}                                               & 0.9954 & \textbf{25.2568}                                            & 22.5873                                            & \textbf{96.91}    \\
                                 & 3DCNN-BVP \cite{yu2019remote}                            & 0.7363                                              & 2.1517                                               & 0.9873 & 26.8545                                            & 25.6097                                           & 96.70    \\ \midrule
\multirow{2}{*}{\begin{tabular}[l]{@{}l@{}}OBF-H \\ (10-folds)\end{tabular}}           & 3DCNN-PEAK                            &0.5824	&1.3059	&\textbf{0.9955}	&25.4702	&\textbf{21.9592}	&96.79    \\
                                 & 3DCNN-BVP \cite{yu2019remote}                            & 1.2912	&3.4956	&0.9676	&33.1952	&37.6494	&95.78    \\ \midrule
\multirow{2}{*}{\begin{tabular}[l]{@{}l@{}}OBF-P \\ (cross dataset)\end{tabular}}           & 3DCNN-PEAK                            & 1.2621	&3.1478&	0.9751	&33.6131	&29.8656	&95.78   \\
                                 & 3DCNN-BVP \cite{yu2019remote}                            & 1.4451	&4.3984	&0.9447	&39.4681	&56.3431	&95.16    \\  \midrule
\multirow{3}{*}{-}               & POS \cite{wang2016algorithmic}                                  &2.1005	&5.3898 &0.9221	&47.0708	&50.3133	&94.13    \\ \cmidrule{2-8}
                                 & CHROM \cite{de2013robust}                                & 4.1686	&7.6021	&0.8434	&93.3947	&51.5717	&88.24    \\ \cmidrule{2-8}
                                 & PBV \cite{de2014improved}                                  & 3.2652 &	6.2210	& 0.9106	& 63.8180	& 52.8818	& 92.24   \\ \bottomrule
\end{tabular}
\begin{tablenotes}
\item MAE = mean absolute error, RMSE = root mean squared error, R = Pearson Correlation, STD = standard deviation.
\end{tablenotes}
\end{threeparttable}
\label{tb:test_OBF_H}
\end{table*}

\begin{table*}[h]
\caption{Results of heart rate and inter-beat interval on the whole OBF dataset}
\centering
\begin{threeparttable}
\begin{tabular}{l|c c c | c c c}
\toprule
\multirow{2}{*}{} & \multicolumn{3}{c|}{Heart Rate}                                                           & \multicolumn{3}{c}{Inter-beat Interval}                                                                     \\ \cmidrule{2-7} 
                  & \begin{tabular}[c]{@{}c@{}}MAE\\ (bpm)\end{tabular} & \begin{tabular}[c]{@{}c@{}}RMSE\\ (bpm)\end{tabular} & R      & \begin{tabular}[c]{@{}c@{}}MAE\\ (ms)\end{tabular} & \begin{tabular}[c]{@{}c@{}}STD\\ (ms)\end{tabular} & Accuracy \\ \midrule
3DCNN-PEAK        & \textbf{1.4658} & \textbf{3.5351} & \textbf{0.9686} & \textbf{50.7481} &\textbf{ 68.0636} & \textbf{94.18}    \\ \midrule
3DCNN-BVP\cite{yu2019remote}         & 2.1371 & 5.0769 & 0.9347 & 58.2923 & 91.6229 & 93.40    \\ \midrule
POS\cite{wang2016algorithmic}        & 4.0234 & 7.6933 & 0.8568 & 81.3753 & 82.3035 & 90.36    \\ \midrule
CHROM\cite{de2013robust}             & 5.6161 & 9.2114 & 0.7900 & 120.6662 & 77.8512 & 85.53    \\ \midrule
PBV\cite{de2014improved}               & 6.0576 & 10.0119 & 0.7711 & 108.2304 & 100.0128 & 87.63    \\ \bottomrule
\end{tabular}
\begin{tablenotes}
\item MAE = mean absolute error, RMSE = root mean squared error, R = Pearson Correlation, STD = standard deviation.
\end{tablenotes}
\end{threeparttable}
\label{tb:test_all}
\end{table*}

\subsection{Results for Experiment II: AF/AFL detection}

Fig. \ref{fig:AF_clip} shows the accuracy, sensitivity, and specificity with clip lengths of 10s, 20s, 30s, 60s, and 120s for both healthy vs. AF and SR vs. AF. In general, ECG has the best performance among all, and accuracy is high even with the shortest 10s clip length. Our proposed 3DCNN-PEAK outperforms other remote PPG methods at most clip lengths. We can observe that longer clips can provide better performance. However, longer recording time also decreases the convenience and application scopes of our method. Therefore, we need to balance the recording time and classification performance. It can be observed that the classification performance increases significantly from 10s to 30s but does not increase too much after 30s for our method and most other methods. In addition, the 30s recording time is not very long. Therefore, we select 30s length to show results from our method and other methods in the following part.

\textcolor{black}{Table \ref{tb:af_health} shows the classification results for healthy vs. AF when the clip length is 30s. The accuracy, sensitivity, and specificity for our method are 96.00\%, 95.36\%, and 96.12\%, respectively. For SR vs. AF results in Table \ref{tb:af_sr}, the accuracy, sensitivity, and specificity for our method are 95.23\%, 98.53\%, and 91.12\%, respectively. For SR vs. AFL in Table \ref{tb:afl_sr}, the accuracy, sensitivity, and specificity for our method are 88.43\%, 90.04\%, and 87.04\%, respectively. Compared with other methods, our method performs the best among remote PPG methods and is close to ECG and contact PPG methods.}

\begin{table}[h]
\caption{Results of classification between healthy subjects and patients with AF for 30s clips}
\centering
\begin{threeparttable}
\fontsize{7.5}{8}\selectfont
\begin{tabular}{l l c c c c c c c}
\toprule

\begin{tabular}[c]{@{}l@{}} Signal \\ types \end{tabular}                  & \begin{tabular}[c]{@{}l@{}} Methods \end{tabular} & \begin{tabular}[c]{@{}l@{}} AC \\ (\%) \end{tabular} & \begin{tabular}[c]{@{}l@{}} SE \\ (\%) \end{tabular} & \begin{tabular}[c]{@{}l@{}} SP \\ (\%) \end{tabular} & \begin{tabular}[c]{@{}l@{}} F1 \\ (\%) \end{tabular} & AUC    \\ \midrule
\multirow{3}{*}{ECG} & Reference         & 99.38         & 99.63            & 99.24 & 99.06 &0.9943   \\
 & Islam et al. \cite{islam2019robust} & 99.27   & 99.24   & 99.28 & 98.87 & 0.9915 \\
 & Fan et al. \cite{fan2018multiscaled} & 97.52   & 99.15    & 96.44 & 97.12 & 0.9806 \\\midrule
\begin{tabular}[c]{@{}l@{}} Contact \\ PPG \end{tabular} & Reference     & 98.94 & 99.08 & 98.86 & 98.22 & 0.9897\\ \midrule
\multirow{5}{*}{\begin{tabular}[c]{@{}l@{}} Remote \\ PPG \end{tabular}} & 3DCNN-PEAK         & \textbf{96.00}         & \textbf{95.36}            &\textbf{96.12} &\textbf{92.67} &\textbf{0.9574}\\ 
& 3DCNN-BVP \cite{yu2019remote}          & 95.32         & 94.51            & 95.56 &91.64 &0.9504\\ 
& Shi et al. \cite{shi2019atrial}         & 90.53         & 79.88            & 94.51 &81.96&0.8720 \\ 
& POS \cite{wang2016algorithmic}          & 91.21         & 90.08            & 91.61 & 85.19 & 0.9084\\
& PBV \cite{de2014improved}               & 89.75         & 86.92            & 90.69 & 82.23 & 0.8881\\ \bottomrule
\end{tabular}
\begin{tablenotes}
\item ECG = electrocardiography, PPG = photoplethysmography, AF = atrial fibrillation, AC = accuracy, SE = Sensitivity, SP = Specificity, AUC = area under curve.
\end{tablenotes}
\end{threeparttable}
\label{tb:af_health}
\end{table}

\begin{table}[h]
\caption{Results of classification between patients with SR and patients with AF for 30s clips}
\centering
\begin{threeparttable}
\fontsize{7.5}{8}\selectfont
\begin{tabular}{l l c c c c c c c}
\toprule

\begin{tabular}[c]{@{}l@{}} Signal \\ types \end{tabular}                  & \begin{tabular}[c]{@{}l@{}} Methods \end{tabular} & \begin{tabular}[c]{@{}l@{}} AC \\ (\%) \end{tabular} & \begin{tabular}[c]{@{}l@{}} SE \\ (\%) \end{tabular} & \begin{tabular}[c]{@{}l@{}} SP \\ (\%) \end{tabular} & \begin{tabular}[c]{@{}l@{}} F1 \\ (\%) \end{tabular} & AUC   \\ \midrule
\multirow{3}{*}{ECG} & Reference &97.89 & 99.31 & 95.85 & 98.13 & 0.9758\\
 & Islam et al. \cite{islam2019robust} & 98.32 & 99.49 & 96.87 & 98.52 & 0.9762 \\
 & Fan et al. \cite{fan2018multiscaled}&  96.53   &  97.95    & 94.93 & 96.11 & 0.9617  \\\midrule
\begin{tabular}[c]{@{}l@{}} Contact \\ PPG \end{tabular}                                  & Reference     & 95.67         & 98.23            & 92.83 & 96.22 & 0.9553\\ \midrule
\multirow{5}{*}{\begin{tabular}[c]{@{}l@{}} Remote \\ PPG \end{tabular}} & 3DCNN-PEAK         &\textbf{95.23}	&\textbf{98.53}	&\textbf{91.12} & \textbf{95.87}& \textbf{0.9482}\\ 
& 3DCNN-BVP \cite{yu2019remote}          & 94.38	& 98.48	 & 88.95 &95.17 & 0.9380\\ 
& Shi et al. \cite{shi2019atrial}         & 89.01	& 92.15	& 84.87 & 89.87 & 0.8851 \\ 
& POS \cite{wang2016algorithmic}               & 92.87 &	97.18 &	87.93 & 93.66 & 0.9256\\
& PBV \cite{de2014improved}               & 88.51	& 95.77	& 79.86 & 90.08 & 0.8782\\ \bottomrule
\end{tabular}
\begin{tablenotes}
\item ECG = electrocardiography, PPG = photoplethysmography. SR = sinus rhythm. AF = atrial fibrillation, AC = accuracy, SE = Sensitivity, SP = Specificity, AUC = area under curve.
\end{tablenotes}
\end{threeparttable}
\label{tb:af_sr}
\end{table}

\begin{table}[h]

\caption{Results of classification between patients with SR and patients with AFL for 30s clips}
\centering
\begin{threeparttable}
\fontsize{7.5}{8}\selectfont
\begin{tabular}{l l c c c c c c c}
\toprule

\begin{tabular}[c]{@{}l@{}} Signal \\ types \end{tabular}                  & \begin{tabular}[c]{@{}l@{}} Methods \end{tabular} & \begin{tabular}[c]{@{}l@{}} AC \\ (\%) \end{tabular} & \begin{tabular}[c]{@{}l@{}} SE \\ (\%) \end{tabular} & \begin{tabular}[c]{@{}l@{}} SP \\ (\%) \end{tabular} & \begin{tabular}[c]{@{}l@{}} F1 \\ (\%) \end{tabular} & AUC   \\ \midrule

ECG& Reference & 91.96 & 92.31 & 91.36 & 91.75 & 0.9183\\ \midrule
\begin{tabular}[c]{@{}l@{}} Contact \\ PPG \end{tabular}                           & Reference & 91.33 & 89.79 & 94.67 & 89.50 & 0.9223 \\ \midrule
\multirow{4}{*}{\begin{tabular}[c]{@{}l@{}} Remote \\ PPG \end{tabular}} & 3DCNN-PEAK &\textbf{88.43} &\textbf{90.04} &\textbf{87.04} &\textbf{88.78} & \textbf{0.8854}\\ 
& 3DCNN-BVP \cite{yu2019remote} & 83.18	& 84.35 & 82.22 & 83.47 & 0.8329\\ 
& Shi et al. \cite{shi2019atrial} & 79.39 & 78.64 & 77.39 & 76.61 & 0.7801\\ 
& POS \cite{wang2016algorithmic} & 77.76 & 64.95 & 79.33 & 67.34 & 0.7214\\
& PBV \cite{de2014improved} & 73.68	& 64.36	& 83.46 &66.98 & 0.7391 \\ \bottomrule
\end{tabular}
\begin{tablenotes}
\item ECG = electrocardiography, PPG = photoplethysmography. SR = sinus rhythm. AFL = atrial flutter, AC = accuracy, SE = Sensitivity, SP = Specificity, AUC = area under curve.
\end{tablenotes}
\end{threeparttable}
\label{tb:afl_sr}
\end{table}

\subsection{Results for Computational Speed}
\textcolor{black}{
Our experiments were conducted on an Intel Xeon E5-2650 2.30 GHz CPU and  Nvidia Tesla V100 GPU. The cropped face is first extracted from the original face videos by OpenFace \cite{baltrusaitis2018openface}. The running speed of this step is 30 fps. We train our 3DCNN model with the whole OBF dataset (about 31 hours of face videos) for 45 epochs, which costs about 45 hours. The inference time of the model is about 2 ms ($9 \times 10^5$ fps) for one 30s video clip. The training of SVM classification takes 60 ms. For each 30s clip, the inference time of SVM classification is about 1.2 ms ($1.5 \times 10^6$ fps). 
}

\subsection{Discussion}

\textcolor{black}{From experiment I: remote PPG monitoring, there are three points we can conclude from Table \ref{tb:test_OBF_P}, \ref{tb:test_OBF_H}, and \ref{tb:test_all}. 1) The evaluation of OBF-P, OBF-H, and the whole OBF dataset shows that our proposed method can achieve the best performance of remote PPG monitoring on both healthy subject data and patient data. 2) Patient data is more challenging than healthy subject data, but the proposed 3DCNN-PEAK can achieve better results than other baselines on patient data. The results demonstrate that the proposed method with learning systolic peaks effectively improves remote PPG monitoring performance for patients. 3) Training with the whole OBF dataset can perform better than training with only OBF-P or OBF-H. Therefore, we use the remote PPG signals from the model trained on the whole OBF dataset for our classification tasks.}

\textcolor{black}{From experiment II: AF/AFL detection, two points can be observed from the results. First, the ECG and contact PPG achieve high accuracy, which validates that the HRV features are effective for AF/AFL detection. Second, our proposed 3DCNN-PEAK achieves slightly lower performance than the contact PPG, but it works best among all remote PPG methods, which indicates that learning systolic peaks can facilitate the AF/AFL detection.}

\textcolor{black}{We also discuss the performance difference among healthy vs. AF, SR vs. AF, and SR vs. AFL. It can be seen that the results of SR vs. AF are slightly lower than that of healthy vs. AF, which means the SR vs. AF classification task is more challenging. This is expected since both AF and SR samples are from the patients, while the classification of AF vs. healthy uses data from two different groups of subjects whose data is more heterogeneous. In addition, results in Experiment I show that patient data is more challenging and has lower remote PPG monitoring performance, which may cause low classification performance between SR patients and AF patients. It is noted that the performance of all methods for SR vs. AFL is lower than healthy vs. AF or SR vs. AF. There are two reasons. 1) There is only a limited number of AFL patients in our dataset, and the numbers of SR (61 subjects) and AFL (11 subjects) are highly unbalanced, which might cause the performance decreasing. A larger AFL dataset could give a performance improvement and more reliable evaluation. 2) It was reported that adding PPG morphology features helped detect AFL compared to only using HRV features \cite{eerikainen2019detecting}. Since only HRV features are used in our experiment, future works could focus on extracting the remote PPG morphology from face videos to improve AFL detection.}
 
\textcolor{black}{We checked some cases when the detector fails. One reason is that some facial videos contain spontaneous facial motions that lead to noises in remote PPG. Another reason is that some PPG signals from patients contain systolic peaks with very low amplitudes due to small blood stroke volume \cite{valiaho2019wrist}. The two factors may cause wrongly detected peaks and eventually lead to wrong classification.}

\textcolor{black}{Our study may lead to healthcare products, e.g., self-screening of AF symptoms of suspectable populations at home or self-monitoring for the chronic patients to check for recurrence of AF after treatment. Non-contact AF detection can also be used in public places for rapids screening of AF since this method does not need time to attach sensors to the subject, and the computational speed is high enough for real-time application. In addition, the non-contact way is more hygienic without contact sensors.}

\section{Conclusion}

\textcolor{black}{In this paper, we propose a novel method for non-contact AF detection by utilizing the systolic peaks to train our deep learning model. A large-scale OBF dataset was collected for non-contact AF detection from face videos. We achieve high accuracy for non-contact AF detection and demonstrate the feasibility of non-contact AFL detection.}

\section*{Acknowledgements}
\textcolor{black}{The authors would like to thank Jingang Shi, Iman Alikhani, and Zitong Yu from University of Oulu for their important contributions in data collection. The authors wish to acknowledge CSC – IT Center for Science, Finland, for computational resources.}

\textcolor{black}{
\bibliographystyle{ieeetr}
\bibliography{refs}

\begin{thebibliography}{10}

\bibitem{authors20122012}
A.~F. Members, A.~J. Camm, G.~Y. Lip, R.~De~Caterina, I.~Savelieva, D.~Atar,
  S.~H. Hohnloser, G.~Hindricks, P.~Kirchhof, E.~C. for Practice
  Guidelines~(CPG), {\em et~al.}, ``2012 focused update of the esc guidelines
  for the management of atrial fibrillation: an update of the 2010 esc
  guidelines for the management of atrial fibrillation developed with the
  special contribution of the european heart rhythm association,'' {\em
  European heart journal}, vol.~33, no.~21, pp.~2719--2747, 2012.

\bibitem{fukunami1991detection}
M.~Fukunami, T.~Yamada, M.~Ohmori, K.~Kumagai, K.~Umemoto, A.~Sakai, N.~Kondoh,
  T.~Minamino, and N.~Hoki, ``Detection of patients at risk for paroxysmal
  atrial fibrillation during sinus rhythm by p wave-triggered signal-averaged
  electrocardiogram.,'' {\em Circulation}, vol.~83, no.~1, pp.~162--169, 1991.

\bibitem{mehta2009detection}
S.~Mehta, N.~Lingayat, and S.~Sanghvi, ``Detection and delineation of p and t
  waves in 12-lead electrocardiograms,'' {\em Expert Systems}, vol.~26, no.~1,
  pp.~125--143, 2009.

\bibitem{guidera1993signal}
S.~A. Guidera and J.~S. Steinberg, ``The signal-averaged p wave duration: a
  rapid and noninvasive marker of risk of atrial fibrillation,'' {\em Journal
  of the American College of Cardiology}, vol.~21, no.~7, pp.~1645--1651, 1993.

\bibitem{couceiro2008detection}
R.~Couceiro, P.~Carvalho, J.~Henriques, M.~Antunes, M.~Harris, and J.~Habetha,
  ``Detection of atrial fibrillation using model-based ecg analysis,'' in {\em
  2008 19th International conference on pattern recognition}, pp.~1--5, IEEE,
  2008.

\bibitem{tateno2001automatic}
K.~Tateno and L.~Glass, ``Automatic detection of atrial fibrillation using the
  coefficient of variation and density histograms of rr and $\delta$rr
  intervals,'' {\em Medical and Biological Engineering and Computing}, vol.~39,
  no.~6, pp.~664--671, 2001.

\bibitem{dash2009automatic}
S.~Dash, K.~Chon, S.~Lu, and E.~Raeder, ``Automatic real time detection of
  atrial fibrillation,'' {\em Annals of biomedical engineering}, vol.~37,
  no.~9, pp.~1701--1709, 2009.

\bibitem{islam2016rhythm}
M.~S. Islam, N.~Ammour, N.~Alajlan, and H.~Aboalsamh, ``Rhythm-based heartbeat
  duration normalization for atrial fibrillation detection,'' {\em Computers in
  biology and medicine}, vol.~72, pp.~160--169, 2016.

\bibitem{islam2019robust}
M.~S. Islam, M.~M.~B. Ismail, O.~Bchir, M.~Zakariah, and Y.~A. Alotaibi,
  ``Robust detection of atrial fibrillation using classification of a
  linearly-transformed window of rr intervals tachogram,'' {\em IEEE Access},
  vol.~7, pp.~110012--110022, 2019.

\bibitem{shashikumar2017deep}
S.~P. Shashikumar, A.~J. Shah, Q.~Li, G.~D. Clifford, and S.~Nemati, ``A deep
  learning approach to monitoring and detecting atrial fibrillation using
  wearable technology,'' in {\em 2017 IEEE EMBS international conference on
  biomedical \& health informatics (BHI)}, pp.~141--144, IEEE, 2017.

\bibitem{fan2018multiscaled}
X.~Fan, Q.~Yao, Y.~Cai, F.~Miao, F.~Sun, and Y.~Li, ``Multiscaled fusion of
  deep convolutional neural networks for screening atrial fibrillation from
  single lead short ecg recordings,'' {\em IEEE journal of biomedical and
  health informatics}, vol.~22, no.~6, pp.~1744--1753, 2018.

\bibitem{pourbabaee2018deep}
B.~Pourbabaee, M.~J. Roshtkhari, and K.~Khorasani, ``Deep convolutional neural
  networks and learning ecg features for screening paroxysmal atrial
  fibrillation patients,'' {\em IEEE Transactions on Systems, Man, and
  Cybernetics: Systems}, vol.~48, no.~12, pp.~2095--2104, 2018.

\bibitem{poh2018diagnostic}
M.-Z. Poh, Y.~C. Poh, P.-H. Chan, C.-K. Wong, L.~Pun, W.~W.-C. Leung, Y.-F.
  Wong, M.~M.-Y. Wong, D.~W.-S. Chu, and C.-W. Siu, ``Diagnostic assessment of
  a deep learning system for detecting atrial fibrillation in pulse
  waveforms,'' {\em Heart}, vol.~104, no.~23, pp.~1921--1928, 2018.

\bibitem{kwon2019deep}
S.~Kwon, J.~Hong, E.-K. Choi, E.~Lee, D.~E. Hostallero, W.~J. Kang, B.~Lee,
  E.-R. Jeong, B.-K. Koo, S.~Oh, {\em et~al.}, ``Deep learning approaches to
  detect atrial fibrillation using photoplethysmographic signals: algorithms
  development study,'' {\em JMIR mHealth and uHealth}, vol.~7, no.~6,
  p.~e12770, 2019.

\bibitem{eerikainen2019detecting}
L.~M. Eerik{\"a}inen, A.~G. Bonomi, F.~Schipper, L.~R. Dekker, H.~M. de~Morree,
  R.~Vullings, and R.~M. Aarts, ``Detecting atrial fibrillation and atrial
  flutter in daily life using photoplethysmography data,'' {\em IEEE journal of
  biomedical and health informatics}, vol.~24, no.~6, pp.~1610--1618, 2019.

\bibitem{lee2012atrial}
J.~Lee, B.~A. Reyes, D.~D. McManus, O.~Maitas, and K.~H. Chon, ``Atrial
  fibrillation detection using an iphone 4s,'' {\em IEEE Transactions on
  Biomedical Engineering}, vol.~60, no.~1, pp.~203--206, 2012.

\bibitem{krivoshei2017smart}
L.~Krivoshei, S.~Weber, T.~Burkard, A.~Maseli, N.~Brasier, M.~K{\"u}hne,
  D.~Conen, T.~Huebner, A.~Seeck, and J.~Eckstein, ``Smart detection of atrial
  fibrillation,'' {\em EP Europace}, vol.~19, no.~5, pp.~753--757, 2017.

\bibitem{bonomi2018atrial}
A.~G. Bonomi, F.~Schipper, L.~M. Eerik{\"a}inen, J.~Margarito, R.~Van~Dinther,
  G.~Muesch, H.~M. De~Morree, R.~M. Aarts, S.~Babaeizadeh, D.~D. McManus, {\em
  et~al.}, ``Atrial fibrillation detection using a novel cardiac ambulatory
  monitor based on photo-plethysmography at the wrist,'' {\em Journal of the
  American Heart Association}, vol.~7, no.~15, p.~e009351, 2018.

\bibitem{dorr2019watch}
M.~D{\"o}rr, V.~Nohturfft, N.~Brasier, E.~Bosshard, A.~Djurdjevic, S.~Gross,
  C.~J. Raichle, M.~Rhinisperger, R.~St{\"o}ckli, and J.~Eckstein, ``The watch
  af trial: Smartwatches for detection of atrial fibrillation,'' {\em JACC:
  Clinical Electrophysiology}, vol.~5, no.~2, pp.~199--208, 2019.

\bibitem{corino2017detection}
V.~D. Corino, R.~Laureanti, L.~Ferranti, G.~Scarpini, F.~Lombardi, and L.~T.
  Mainardi, ``Detection of atrial fibrillation episodes using a wristband
  device,'' {\em Physiological measurement}, vol.~38, no.~5, p.~787, 2017.

\bibitem{valiaho2019wrist}
E.-S. V{\"a}liaho, P.~Kuoppa, J.~Lipponen, T.~Martikainen, H.~J{\"a}ntti,
  T.~Rissanen, I.~Kolk, M.~Castr{\'e}n, J.~Halonen, M.~Tarvainen, {\em et~al.},
  ``Wrist band photoplethysmography in detection of individual pulses in atrial
  fibrillation and algorithm-based detection of atrial fibrillation,'' {\em EP
  Europace}, vol.~21, no.~7, pp.~1031--1038, 2019.

\bibitem{wu2000photoplethysmography}
T.~Wu, V.~Blazek, and H.~J. Schmitt, ``Photoplethysmography imaging: a new
  noninvasive and noncontact method for mapping of the dermal perfusion
  changes,'' in {\em Optical techniques and instrumentation for the measurement
  of blood composition, structure, and dynamics}, vol.~4163, pp.~62--70,
  International Society for Optics and Photonics, 2000.

\bibitem{iozzia2016relationships}
L.~Iozzia, L.~Cerina, and L.~Mainardi, ``Relationships between heart-rate
  variability and pulse-rate variability obtained from video-ppg signal using
  zca,'' {\em Physiological measurement}, vol.~37, no.~11, p.~1934, 2016.

\bibitem{couderc2015detection}
J.-P. Couderc, S.~Kyal, L.~K. Mestha, B.~Xu, D.~R. Peterson, X.~Xia, and
  B.~Hall, ``Detection of atrial fibrillation using contactless facial video
  monitoring,'' {\em Heart Rhythm}, vol.~12, no.~1, pp.~195--201, 2015.

\bibitem{verkruysse2008remote}
W.~Verkruysse, L.~O. Svaasand, and J.~S. Nelson, ``Remote plethysmographic
  imaging using ambient light.,'' {\em Optics express}, vol.~16, no.~26,
  pp.~21434--21445, 2008.

\bibitem{us_burn}
``Burn incidence fact sheet,'' {\em American Burn Association}.
\newblock https://ameriburn.org/who-we-are/media/burn-incidence-fact-sheet/.

\bibitem{li2018obf}
X.~Li, I.~Alikhani, J.~Shi, T.~Seppanen, J.~Junttila, K.~Majamaa-Voltti,
  M.~Tulppo, and G.~Zhao, ``The obf database: A large face video database for
  remote physiological signal measurement and atrial fibrillation detection,''
  in {\em 2018 13th IEEE International Conference on Automatic Face \& Gesture
  Recognition (FG 2018)}, pp.~242--249, IEEE, 2018.

\bibitem{yu2019remote}
Z.~Yu, X.~Li, and G.~Zhao, ``Remote photoplethysmograph signal measurement from
  facial videos using spatio-temporal networks,'' in {\em 30th British Machine
  Vision Conference 2019, {BMVC} 2019, Cardiff, UK, September 9-12, 2019},
  p.~277, {BMVA} Press, 2019.

\bibitem{shi2019atrial}
J.~Shi, I.~Alikhani, X.~Li, Z.~Yu, T.~Sepp{\"a}nen, and G.~Zhao, ``Atrial
  fibrillation detection from face videos by fusing subtle variations,'' {\em
  IEEE Transactions on Circuits and Systems for Video Technology}, vol.~30,
  no.~8, pp.~2781--2795, 2019.

\bibitem{yan2018contact}
B.~P. Yan, W.~H. Lai, C.~K. Chan, S.~C.-H. Chan, L.-H. Chan, K.-M. Lam, H.-W.
  Lau, C.-M. Ng, L.-Y. Tai, K.-W. Yip, {\em et~al.}, ``Contact-free screening
  of atrial fibrillation by a smartphone using facial pulsatile
  photoplethysmographic signals,'' {\em Journal of the American Heart
  Association}, vol.~7, no.~8, p.~e008585, 2018.

\bibitem{corino2020simple}
V.~D. Corino, L.~Iozzia, G.~Scarpini, L.~T. Mainardi, and F.~Lombardi, ``A
  simple model to detect atrial fibrillation via visual imaging,'' {\em
  Biomedical Engineering/Biomedizinische Technik}, vol.~65, no.~6,
  pp.~721--728, 2020.

\bibitem{islam2017atrial}
S.~Islam, N.~Ammour, and N.~Alajlan, ``Atrial fibrillation detection with
  multiparametric rr interval feature and machine learning technique,'' in {\em
  2017 International Conference on Informatics, Health \& Technology (ICIHT)},
  pp.~1--5, IEEE, 2017.

\bibitem{andersen2019deep}
R.~S. Andersen, A.~Peimankar, and S.~Puthusserypady, ``A deep learning approach
  for real-time detection of atrial fibrillation,'' {\em Expert Systems with
  Applications}, vol.~115, pp.~465--473, 2019.

\bibitem{butkuviene2021considerations}
M.~Butkuviene, A.~Petrenas, A.~Solosenko, A.~Martin-Yebra, V.~Marozas, and
  L.~Sornmo, ``Considerations on performance evaluation of atrial fibrillation
  detectors,'' {\em IEEE Transactions on Biomedical Engineering}, 2021.

\bibitem{bruser2012automatic}
C.~Bruser, J.~Diesel, M.~D. Zink, S.~Winter, P.~Schauerte, and S.~Leonhardt,
  ``Automatic detection of atrial fibrillation in cardiac vibration signals,''
  {\em IEEE journal of biomedical and health informatics}, vol.~17, no.~1,
  pp.~162--171, 2012.

\bibitem{hurnanen2016automated}
T.~Hurnanen, E.~Lehtonen, M.~J. Tadi, T.~Kuusela, T.~Kiviniemi, A.~Saraste,
  T.~Vasankari, J.~Airaksinen, T.~Koivisto, and M.~P{\"a}nk{\"a}{\"a}l{\"a},
  ``Automated detection of atrial fibrillation based on time--frequency
  analysis of seismocardiograms,'' {\em IEEE journal of biomedical and health
  informatics}, vol.~21, no.~5, pp.~1233--1241, 2016.

\bibitem{lahdenoja2017atrial}
O.~Lahdenoja, T.~Hurnanen, Z.~Iftikhar, S.~Nieminen, T.~Knuutila, A.~Saraste,
  T.~Kiviniemi, T.~Vasankari, J.~Airaksinen, M.~P{\"a}nk{\"a}{\"a}l{\"a}, {\em
  et~al.}, ``Atrial fibrillation detection via accelerometer and gyroscope of a
  smartphone,'' {\em IEEE Journal of Biomedical and Health Informatics},
  vol.~22, no.~1, pp.~108--118, 2017.

\bibitem{wang2016algorithmic}
W.~Wang, A.~C. den Brinker, S.~Stuijk, and G.~De~Haan, ``Algorithmic principles
  of remote ppg,'' {\em IEEE Transactions on Biomedical Engineering}, vol.~64,
  no.~7, pp.~1479--1491, 2016.

\bibitem{feng2014motion}
L.~Feng, L.-M. Po, X.~Xu, Y.~Li, and R.~Ma, ``Motion-resistant remote imaging
  photoplethysmography based on the optical properties of skin,'' {\em IEEE
  Transactions on Circuits and Systems for Video Technology}, vol.~25, no.~5,
  pp.~879--891, 2014.

\bibitem{baltrusaitis2018openface}
T.~Baltrusaitis, A.~Zadeh, Y.~C. Lim, and L.-P. Morency, ``Openface 2.0: Facial
  behavior analysis toolkit,'' in {\em 2018 13th IEEE international conference
  on automatic face \& gesture recognition (FG 2018)}, pp.~59--66, IEEE, 2018.

\bibitem{mcduff2014remote}
D.~McDuff, S.~Gontarek, and R.~W. Picard, ``Remote detection of
  photoplethysmographic systolic and diastolic peaks using a digital camera,''
  {\em IEEE Transactions on Biomedical Engineering}, vol.~61, no.~12,
  pp.~2948--2954, 2014.

\bibitem{cazelles2020wasserstein}
E.~Cazelles, A.~Robert, and F.~Tobar, ``The wasserstein-fourier distance for
  stationary time series,'' {\em IEEE Transactions on Signal Processing},
  vol.~69, pp.~709--721, 2020.

\bibitem{makowski2021neurokit2}
D.~Makowski, T.~Pham, Z.~J. Lau, J.~C. Brammer, F.~Lespinasse, H.~Pham,
  C.~Sch{\"o}lzel, and S.~A. Chen, ``Neurokit2: A python toolbox for
  neurophysiological signal processing,'' {\em Behavior Research Methods},
  pp.~1--8, 2021.

\bibitem{de2013robust}
G.~De~Haan and V.~Jeanne, ``Robust pulse rate from chrominance-based rppg,''
  {\em IEEE Transactions on Biomedical Engineering}, vol.~60, no.~10,
  pp.~2878--2886, 2013.

\bibitem{de2014improved}
G.~De~Haan and A.~Van~Leest, ``Improved motion robustness of remote-ppg by
  using the blood volume pulse signature,'' {\em Physiological measurement},
  vol.~35, no.~9, p.~1913, 2014.

\bibitem{liu2020detecting}
X.~Liu, X.~Yang, J.~Jin, and A.~Wong, ``Detecting pulse wave from unstable
  facial videos recorded from consumer-level cameras: A disturbance-adaptive
  orthogonal matching pursuit,'' {\em IEEE Transactions on Biomedical
  Engineering}, vol.~67, no.~12, pp.~3352--3362, 2020.

\end{thebibliography}
}
\end{document}